\documentclass[aap,seceqn,citesort,noautosecdot,MSNbibl,dvips]{arximspdf}
\usepackage{multirow}
\usepackage{graphicx}

\doi{10.1214/09-AAP654}
\volume{20}
\issue{4}
\pubyear{2010}
\firstpage{1425}
\lastpage{1469}

\makeatletter

\newtheorem{theo}{Theorem}[section]
\newproclaim{rem}[theo]{Remark}
\newtheorem{lem}[theo]{Lemma}
\newtheorem{prop}[theo]{Proposition}

\newproclaim{Assumption}{Assumption}

  \let\sv@tabnotetext\tabnotetext
  \let\sv@tabnotemark@fmt\tabnotemark@fmt
   \long\def\legend#1{{\let\tabnote@indent\leavevmode\sv@tabnotetext[]{}{#1}}}

\makeatother

\begin{document}
\begin{frontmatter}

\title{Do price and volatility jump together?}
\runtitle{Do price and volatility jump together?}

\begin{aug}
\author[A]{\fnms{Jean} \snm{Jacod}\ead[label=e1]{jean.jacod@upmc.fr}} and
\author[B]{\fnms{Viktor} \snm{Todorov}\corref{}\ead[label=e2]{v-todorov@kellogg.northwestern.edu}}
\runauthor{J. Jacod and V. Todorov}
\affiliation{UPMC (Universit\'e Paris-6) and Northwestern University}
\address[A]{Institut de Math\'ematiques de Jussieu\\
CNRS---UMR 7586\\
Universit\'e Pierre et
Marie Curie--P6\\
4 Place Jussieu\\
75252 Paris-Cedex 05\\
France\\
\printead{e1}}
\address[B]{Department of Finance\\
Kellogg School of Management\\
Northwestern University\\
Evanston, Illinois 60208-2001\\
USA\\
\printead{e2}}
\end{aug}

\received{\smonth{7} \syear{2009}}
\revised{\smonth{10} \syear{2009}}

%
\begin{abstract}
We consider a process $X_t$, which is observed on a finite time
interval $[0,T]$, at discrete times $0,\Delta_n,2\Delta_n,\ldots.$ This
process is an It\^o semimartingale with stochastic volatility
$\sigma_t^2$. Assuming that $X$ has jumps on $[0,T]$, we derive tests
to decide whether the volatility process has jumps occurring
simultaneously with the jumps of $X_t$. There are two different
families of tests for the two possible null hypotheses (common jumps or
disjoint jumps). They have a prescribed asymptotic level as the mesh
$\Delta_n$ goes to $0$. We show on some simulations that these tests
perform reasonably well even in the finite sample case, and we also put
them in use on S\&P 500 index data.
\end{abstract}

%

\end{frontmatter}

\section{\texorpdfstring{Introduction.}{Introduction}}\label{sec-Intro}

Financial asset prices have two well-documented sa\-lient
features: their volatility changes over time and
their trajectories can exhibit large discontinuities. Both
features have nontrivial implications for risk
modeling and management as the underlying asset itself is no
longer sufficient to span all the available risks in it and
derivatives (written on it) are typically needed. Of central
importance then becomes the relationship between the price jumps
and volatility. For example, if the volatility is driven by a
single (Markov) diffusion process, then one can separate the
management of volatility and jump risks by using first
at-the-money options for the former and then out-of-the-money
options for the latter. But such a simple separate management of
these two risks will obviously not work if the price jumps are
associated with simultaneous discontinuous changes in the level of
volatility. Empirical evidence in \cite{TT} based on the behavior
of close-to-maturity options written on the stock market index
suggest that this indeed might be the case. And this is exactly
what we try to investigate in this paper: are price jumps
accompanied by jumps in volatility?

The link between price and volatility jumps is intrinsically associated with
the observed path, and therefore we
develop tests that are, as much as possible, independent from the underlying
model. More specifically, we suppose that we have discrete observations
from an arbitrary It\^{o} semimartingale (typically the log-price) at times
$i\Delta_n$ for
$i=0,1,\ldots,[T/\Delta_n]$ where the time span $T$ will stay fixed and the
length of the high-frequency intervals $\Delta_n\rightarrow0$. Under such
a sampling scheme, we propose tests that determine the common arrival,
or not, of
the price and volatility jumps on the discretely-observed path over $[0,T]$.

The test statistics that we construct can be intuitively described
as follows. First, we identify the high-frequency price increments
containing jumps as those being higher in absolute value than a
truncation level which goes to zero at a certain (known) rate.
Then, for the set of identified jump times we construct left and
right local volatility estimators from the neighboring
high-frequency price increments. Our statistics are simple sums of
certain functions of the identified jumps and the associated left
and right volatility estimators. Then the tests we develop are
based on the different limit behavior of these statistics on the
sets of common and disjoint arrival of the price and volatility
jumps.

While the results in the paper are derived for general functions
measuring the distance between the left and right volatility,
there is one specific choice which is particularly attractive for
our testing purposes, and we use it in our applications. This
function corresponds to the log-likelihood ratio test for deciding
whether two independent samples of i.i.d. zero-mean normal
variables have the same variance. The link with our analysis comes
from the fact that the leading terms in the asymptotic expansions
of the left and right local volatility estimators are (close to) sample
averages of squared increments of a Brownian motion multiplied
by the volatility level straight before and after the price jump
time. The ``local Gaussianity'' of the high-frequency increments
has been also used in \cite{MZ} in a different context, that is, for
constructing various integrated measures of volatility in a
continuous setting. Unlike \cite{MZ}, however, our analysis is for
processes with jumps.

Finally, our results can be related to \cite{JT} in which we
propose tests for deciding the common arrival of jumps for two
discretely observed processes. The major difference with that
paper is that here one of the processes, namely the volatility, is
not directly observed, and it has to be estimated from the price
increments first. This has nontrivial consequences, as it is
essentially the error associated with measuring the volatility
that determines the asymptotic behavior of our statistics, and it
can significantly slow down their rate of converge. The
intrinsic nonsymmetric nature of the price and volatility is
reflected in our construction of the tests here, and this makes
the statistical problem very different from the one analyzed in
\cite{JT}.

The paper is organized as follows. Section \ref{sec:SA} introduces our setup
and states the assumptions to be used in the rest of the paper. In
Section \ref{sec:KP} we propose statistics constructed from the
high-frequency data to measure the simultaneous arrival of price
and volatility jumps. In this section we also derive central limit
theorems associated with the statistics. Section \ref{sec:TS}
constructs our
tests using the statistics of Section \ref{sec:KP}. Section \ref{sec:MC}
contains Monte Carlo evidence for the performance of the tests, while
Section \ref{sec:DATA} applies our tests to real financial data.
Proofs are in Section~\ref{proofs}.\looseness=1

\section{\texorpdfstring{Setting and assumptions.}{Setting and assumptions}}\label{sec:SA}

We suppose throughout that our underlying process $X$ is
an It\^{o} semimartingale on a filtered space $(\Omega,\mathcal
F,(\mathcal F_t)_{t\geq0},\mathbb{P})$. This means that
it can be written as
%
%
\begin{eqnarray}\label{S1}
X_{t}&=&X_{0}+\int_{0}^{t}b_{s}\,ds+\int_{0}^{t}\sigma_{s}
\,dW_{s}\nonumber\\
&&{} +\int_{0}^{t}\int_E\bigl(\delta(s,z)1_{\{|\delta(t,z)|\leq1\}
}\bigr)(\mu
-\nu)(ds,dz)\\
&&{}
+\int_{0}^{t}\int_E\bigl(\delta(s,z)1_{\{|\delta(t,z)|>1\}}\bigr)\mu
(ds,dz),\nonumber
\end{eqnarray}
where $W$ is a standard Brownian motion, and $\mu$ is a
Poisson random measure on $[0,\infty)\times E$, with $(E,\mathcal E)$ an
auxiliary measurable space, on the space $(\Omega,\mathcal F,(\mathcal
F_t)_{t\geq0},\mathbb{P})$ and the
predictable compensator (or intensity measure) of $\mu$ is $\nu
(ds,dz)=ds\otimes\lambda(dz)$ for some given $\sigma$-finite
measure $\lambda$ on $(E,\mathcal E)$. We write $c_t=(\sigma_t)^2$
for the volatility
process. The processes $b_t$ and $\sigma_t$ should be
progressively measurable and $\delta(\omega,t,z)$ should be a predictable
function on $\Omega\times\mathbb{R}_+\times\mathbb{E}$. We refer
to \cite{JS} for all
unexplained, but classical, notation.

We need some assumptions on $X$, and below $r\in[0,2)$.
\renewcommand{\theAssumption}{(H-$r$)}
\begin{Assumption}\label{asssumHr}
(a) The process $b_{t}$
is locally bounded.

(b) The process $\sigma_t$ is c\`adl\`ag, and neither $\sigma
_t$ nor
$\sigma_{t-}$ vanish.

(c) We have $|\delta(\omega,t,x)|\leq\Gamma_t(\omega
)\gamma(x)$,
for a locally bounded process $\Gamma_t$ and a (nonrandom) function
$\gamma\geq0$ satisfying $\int_{E}(\gamma(x)^r\wedge1) \lambda
(dx)<\infty$.
\end{Assumption}

When $r=2$ this is little more than $X$ being an
It\^o semimartingale, except for the fact that $\sigma_t$ and
$\sigma_{t-}$ do not vanish. When $r<2$ it requires further that the
jumps are $r$-summable, and the bigger $r$ is, the weaker is the
assumption. When (H-$0$) holds, then the jumps of $X$ have finite activity.

Next, we make an assumption on the local behavior of $\sigma_t$. We want
to accommodate two extreme cases: one is when $\sigma_t$ is itself
an It\^o semimartingale (a quite usual assumption for stochastic volatility
models), and one is when it is the sum of finitely many jumps plus a
continuous process having pathwise some H\"older continuity property
such as a fractional Brownian motion. So we present an assumption which
may look complicated but is satisfied by all models used so far and
implies that $\sigma_t$ is c\`adl\`ag. In
this assumption, $v$ is in $(0,1]$, and the bigger it is, the stronger
is the assumption.
\renewcommand{\theAssumption}{(K-$v$)}
\begin{Assumption}\label{asssumKv}
We have $\sigma_t=\Sigma
(Z_t,\overline{Z}
_t)$, where
$\Sigma$ is a $C^1$ function on~$\mathbb{R}^2$, and $Z_t$ and
$\overline{Z}_t$ are two adapted
processes with the following properties:

(a) The process $Z$ is an It\^o semimartingale satisfying
(H-$2$) when $v\leq1/2$ whereas when $v>1/2$ it satisfies (H-$1/v$), and
its continuous martingale part vanishes.

(b) The process $\overline{Z}_t$ satisfies, for some locally
bounded process
$\Gamma'$,
%
%
\begin{equation}\label{S2}
0<s\leq1 \quad\Rightarrow\quad
|\overline{Z}_{t+s}(\omega)-\overline{Z}_t(\omega)|\leq\Gamma
'_{t+s}(\omega) s^v.
\end{equation}
\end{Assumption}

\section{\texorpdfstring{Limit theorems for functionals
of jumps and volatility.}{Limit theorems for functionals of jumps and volatility}}\label{sec:KP}

Our aim is to decide whether we have jumps of $X$ and $c$ occurring at
the same times, and for this we make use of the following processes
where $\Delta Y_t=Y_t-Y_{t-}$ is the jump size at time $t$ of any
c\`adl\`ag process $Y$:
%
%
\begin{equation}\label{Z0}
U(F)_t=\sum_{s\leq t}F(\Delta X_s,c_{s-},c_s) 1_{\{\Delta X_s\neq0\}}.
\end{equation}
Here, $F$ is a function on $\mathbb{R}\times\mathbb{R}_+^*\times
\mathbb{R}^*_+$ where
$\mathbb{R}_+^*=(0,\infty)$. The derivatives
of $F$, when they exist, are denoted by $F'_j$ and $F''_{jk}$, for
$j,k=1,2,3$. The general idea will be to choose a function $F$ which,
for example, is nonnegative and $F(x,y,z)=0$ if and only if $y=z$; then
$U(F)_T>0$ on the set where the two processes $X$ and $c$ have common
jumps within the time interval $[0,T]$, and $U(F)_T=0$ elsewhere.

The process $U(F)$ is not directly observable because we only observe
$X_{i\Delta_n}$ for $i\in\mathbb{N}$. Consequently, we
``approximate'' it by an
observable process which we presently describe. We need some notation.
For any process $Y$ we set
%
%
\begin{equation}\label{Z4}
\Delta^n_iY=Y_{i\Delta_n}-Y_{(i-1)\Delta_n}.
\end{equation}

We choose two sequences $u_n>0$ and $k_n\in\mathbb{N}^*$ which serve as
cutoff level and window size at stage $n$: we must have $u_n\to0$
but more slowly than $\sqrt{\Delta_n}$, and $k_n\to\infty$ but more
slowly than
$1/\Delta_n$. To this end it is convenient to choose two exponents
$\varpi$ and $\rho$ such that, for some constant $K$,
%
%
\begin{equation}\label{Z5}\qquad
\frac1K\leq\frac{u_n}{\Delta_n^\varpi}\leq K,\qquad
\frac1K\leq k_n\Delta_n^\rho\leq K\qquad \mbox{with }
0<\varpi<\frac12,0<\rho<1.
\end{equation}

The next variables serve as ``local estimators'' of the
volatility:
%
%
\begin{equation}\label{Z6}
\widehat{c}(k_n)_i=\frac1{k_n\Delta_n}\sum_{j=1}^{k_n}
|\Delta^n_{i+j}X|^2 1_{\{|\Delta^n_{i+j}X|\leq u_n\}}.
\end{equation}
Note that (b) of Assumption \ref{asssumHr} implies that $\Delta^n_iX\neq0$ a.s. for all $i,n$,
so $\widehat{c}(k_n)_i>0$ a.s. and we can set
%
%
\begin{equation}\label{Z7}
U(F,k_n)_t=\sum_{i=k_n+1}^{[t/\Delta_n]-k_n}F(\Delta^n_iX,\widehat
{c}(k_n)_{i-k_n-1},\widehat{c}(k_n)_i) 1_{\{|\Delta^n_i
X|>u_n\}}.
\end{equation}
The aim of this section is to describe the asymptotic behavior of
those observable processes $U(F,k_n)$.

\subsection{\texorpdfstring{The law of large numbers.}{The law of large numbers}}
Here we describe under which conditions on $F$ we have
$U(F,k_n)\to U(F)$. Basically, this requires
that $F$ be continuous, plus some additional conditions. However,
we want to apply the result when, for example, $F$ has the form
$F(x,y,z)=1_{\{|x|>a\}} g(y,z)$, where $a>0$, and such an $F$ is
of course not continuous: so the desired convergence does
not take place, unless with probability $1$ there is no jump of
$X$ with size $a$ or $-a$. This is why we introduce the following
family $\mathcal R$ of subsets $R$:
%
%
\begin{equation}\label{Z701}
R\in\mathcal R\quad\Leftrightarrow\quad
\begin{array}{l}
\bullet\ \mbox{$R$ is open, with a finite complement;}\\
\bullet\ D=\{x:
\mathbb{P}(\exists s>0\mbox{ with }\Delta X_s=x )>0\}\subset R.
\end{array}
\end{equation}
\begin{theo}\label{TKP-1} Assume Assumption \ref{asssumHr} for some $r<2$ and Assumption \ref{asssumKv} and
(\ref{Z5}), and let $F$ be a Borel function on $\mathbb{R}\times
\mathbb{R}_+^{*2}$
which is continuous at each point of $R\times\mathbb{R}_+^{*2}$ for some
$R\in\mathcal R$. The processes $U(F,k_n)$ converge
in probability, for the Skorokhod topology, to $U(F)$,
as soon as one of the following three sets of hypotheses is satisfied:

\textup{(a)} $F(x,y,z)=0$ for $|x|\leq\varepsilon$ for some $\varepsilon>0$;

\textup{(b)} we have $r=0$;

\textup{(c)} we have $|F(x,y,z)|\leq K|x|^r(1+y+z)$ if $|x|\leq\varepsilon$
for some $\varepsilon,K>0$.
\end{theo}

\subsection{\texorpdfstring{The central limit theorems.}{The central limit theorems}}

The above consistency result is not enough for us, and we need a
central limit theorem (CLT) associated with it. Moreover, in view of
the statistical applications given later, we need a joint CLT for
the process $U(F,k_n)$ and for the similar process $U(F,wk_n)$
obtained by substituting $k_n$ with $wk_n$ for some integer $w\geq2$.

The test function $F$ should satisfy some smoothness conditions in connection
with the index $r$ in Assumption \ref{asssumHr} and involves another index $p\geq1$ as well.
Namely, we suppose that there exist $R\in\mathcal R$ and $\varepsilon
\geq0$ such that:
%
%
\begin{eqnarray}\label{Z-101}\qquad
&&
\bullet\mbox{ $F$}\qquad\mbox{is $C^1$ on }R\times\mathbb{R}_+^{*2};\nonumber
\\
&&
\bullet\ \frac1{|x|^{p-1}} F'_1(x,y,z)\qquad \mbox{is locally bounded on }
R\times\mathbb{R}_+^{*2};\\
&&
\bullet\ \frac1{|x|^r} F'_2(x,y,z),\qquad
\frac1{|x|^r} F'_3(x,y,z) \mbox{ are bounded on }[-\varepsilon
,\varepsilon]\times
\mathbb{R}_+^{*2}\nonumber
\end{eqnarray}
(recall that any $R\in\mathcal R$ contains $[-\varepsilon,\varepsilon
]$ for some $\varepsilon>0$).
When $\varepsilon=0$ the last condition is empty. When $p=1$ the
second condition
is empty.

We need some additional notation. Let $(\Omega',\mathcal F',\mathbb
{P}')$ be an
auxiliary space
endowed with four sequences $(V_p^-)$, $(V_p^+)$,
$(V'^-_p)$ and $(V'^+_p)$ of independent $\mathcal N(0,1)$ variables. We
introduce the following extension $(\widetilde{\Omega},\widetilde
{\mathcal F},\widetilde{\mathbb{P}})$ of $(\Omega,\mathcal F,\mathbb{P})$:
\[
\widetilde{\Omega}=\Omega\times\Omega',\qquad \widetilde{\mathcal
F}=\mathcal F\otimes\mathcal F',\qquad
\widetilde{\mathbb{P}}=\mathbb{P}\otimes\mathbb{P}'.
\]
Any variable or process defined on $\Omega$ or $\Omega'$ will be extended
to $\widetilde{\Omega}$ in the usual way, without change of notation.
We consider an arbitrary sequence $(T_p)_{p\geq1}$ of positive
stopping times on $(\Omega,\mathcal F,(\mathcal F_t)_{t\geq0},\mathbb
{P})$ which exhausts the jumps of $X$: this means
that $T_p\neq T_q$ if $T_p<\infty$ and $q\neq p$, and that for each
$\omega$
the set $\{t\dvtx\Delta X_t\neq0\}$ is contained in $\{T_p\dvtx p\geq1\}$.

Below we assume Assumption \ref{asssumHr}, and $F$ satisfies (\ref{Z-101}).
Then the formulas
%
%
\begin{equation}\label{Z-12}
\cases{
\displaystyle\mathcal U_t=\sum_{p\geq1} \bigl(F'_2(\Delta X_{T_p},c_{T_p-},c_{T_p})
c_{T_p-}\sqrt{2} V^-_p\cr
\displaystyle\hspace*{41.7pt}{} + F'_3(\Delta X_{T_p},c_{T_p-},c_{T_p})
c_{T_p}\sqrt{2} V^+_p \bigr) 1_{\{T_p\leq t\}},\cr
\displaystyle\mathcal U'_t=\sum_{p\geq1} \bigl(F'_2(\Delta X_{T_p},c_{T_p-},c_{T_p})
c_{T_p-}\sqrt{2} V'^-_p\cr
\displaystyle\hspace*{41.7pt}{} + F'_3(\Delta X_{T_p},c_{T_p-},c_{T_p})c_{T_p}\sqrt{2}
V'^+_p \bigr)
1_{\{T_p\leq t\}},}
\end{equation}
define\vspace*{1pt} two c\`adl\`ag adapted processes $\mathcal U$ and $\mathcal U'$
on the extended
filtered
space $(\widetilde{\Omega},\widetilde{\mathcal F},(\widetilde
{\mathcal F}_t)_{t\geq0},\widetilde{\mathbb{P}})$ where $(\widetilde
{\mathcal F}_t)$ is the smallest
filtration which
contains $(\mathcal F_t)$ and such that the variables
$V^+_p,V^-_p,V'^+_p,V'^-_p$
are $\widetilde{\mathcal F}_{T_p}$-measurable. Moreover, conditionally
on $\mathcal F$, these two
processes are independent, with the same (conditional) laws, and
are centered Gaussian martingales (hence with independent increments)
and with the conditional variances
%
\begin{eqnarray}\label{Z-8}
\widetilde{\mathbb{E}}((\mathcal U_t)^2\mid\mathcal F)=\widetilde
{\mathbb{E}}((\mathcal U'_t)^2\mid\mathcal F)=B(F)_t\hspace*{100pt}
\nonumber\\[-8pt]\\[-8pt]
\eqntext{\mbox{where }\displaystyle B(F)_t=2\sum_{s\leq t} \bigl(c_{s-}^2 F'_2(\Delta X_s,c_{s-},c_s)^2
+c_{s}^2 F'_3(\Delta X_s,c_{s-},c_s)^2 \bigr).}
\end{eqnarray}
Moreover, if we modify the exhausting sequence $(T_p)$ we accordingly modify
$\mathcal U_t$ and $\mathcal U'_t$, but we \textit{do not change} their
$\mathcal F$-conditional
laws which is the only relevant property of $(\mathcal U,\mathcal U')$
for the stable
convergence in law below (all these facts are proved, in a slightly
different form, in \cite{JSEMSTAT}; we refer to \cite{JS} for the stable
convergence
in law).
\begin{theo}\label{TKP-2} Assume Assumption \ref{asssumHr} for some $r<2$
and Assumption~\mbox{\ref{asssumKv}} and
(\ref{Z5}) with
%
%
\begin{equation}\label{Z-129}
\rho<\bigl(2\varpi(2-r)\bigr)\wedge\frac{2v}{1+2v}.
\end{equation}
Let $F$ satisfy (\ref{Z-101}) with $\varepsilon\geq0$ when $r=0$ and
$\varepsilon
>0$ otherwise,
and let $w\geq2$ be an integer.

\begin{longlist}
\item If either $r=0$, or $F(x,y,z)=0$ for $|x|\leq\varepsilon$ for
some $\varepsilon
>0$, the
two-dimensional processes
%
%
\begin{equation}\label{Z-11}
\bigl(\sqrt{k_n} \bigl(U(F,k_n)_t-U(F)_t \bigr),
\sqrt{k_n} \bigl(U(F,wk_n)_t-U(F)_t \bigr) \bigr)
\end{equation}
converge stably in law to the process
$ (\mathcal U,\frac1w (\mathcal U+\sqrt{w-1} \mathcal U'))$ in the
Skorokhod sense.

\item Assume that $r>0$, that $F(0,y,z)=0$ and that $p>1+r/2$ in
(\ref{Z-101}). Assume also that $\rho$ and $\varpi$ satisfy
%
%
\begin{equation}\label{Z-103}
\varpi<\frac1{2r},\qquad
\rho<\bigl(2\varpi(p\wedge2-r)\bigr)\wedge\frac{2p-2-r}r\wedge\frac{2v}{1+2v}
\end{equation}
[which is stronger than (\ref{Z-129})].
Then for any fixed $t>0$ the variables (\ref{Z-11}) converge stably in
law to the variables $ (\mathcal U_t,\frac1w (\mathcal U_t+\sqrt{w-1}
\mathcal U'_t))$.
\end{longlist}
\end{theo}

In (ii) above we do not state the ``functional convergence''
(stably in law), although it is probably true. For the
tests we are after in the paper, we need only the finite-dimensional
convergence of the above theorem.

Our second CLT is about the case when the limiting process in the
first CLT vanishes. Another normalization is then needed, and also
stronger smoothness assumptions on $F$. Namely, we assume (\ref{Z-101})
and
%
%
\begin{eqnarray}\label{Z-102}\quad
&&\bullet\ F(x,y,z) \qquad\mbox{is $C^1$ in $x$ and $C^2$ in $(y,z)$
on } R\times\mathbb{R}_+^{*2},\nonumber\\[-8pt]\\[-8pt]
&&\bullet\ \frac1{|x|^r} F''_{ij}(x,y,z)\qquad
\mbox{for $i,j=2,3$ is bounded on }[-\varepsilon,\varepsilon]\times
\mathbb{R}_+^{*2}.\nonumber
\end{eqnarray}
Of course, the limit in Theorem \ref{TKP-2} may vanish under various
circumstances, but for us it is enough to consider the rather simple
situation where there is a Borel set $A\subset\mathbb{R}$ and some
$\eta>0$
such that
%
%
\begin{eqnarray}\label{Z-104}\qquad
&&\bullet\ \mbox{either $[-\eta,\eta]\subset A$}\quad\mbox{or}\quad\mbox{$[-\eta,\eta
]\cap A
=\varnothing$},\nonumber\\
&&\bullet\ x\in A,y\in\mathbb{R}^*_+\quad\Rightarrow\quad
F(x,y,y)=F'_2(x,y,y)=F'_3(x,y,y)=0,\\
&&\bullet\ x\notin A, y,z\in\mathbb{R}_+^* \quad\Rightarrow\quad
F(x,y,z)=0.\nonumber
\end{eqnarray}
Then obviously $U(F)_t=\mathcal U_t=\mathcal U'_t=0$ on the set $\Omega
_t^A$ on
which, for
all $s\leq t$, we have $\Delta\sigma_s=0$ whenever $\Delta X_s\in A
\setminus\{0\}$. When $A=\mathbb{R}$ the set $\Omega^A_t$ is the set
where $X$ and
$\sigma$ have no common jumps on $[0,t]$.

When $F$ satisfies (\ref{Z-102}), and with a given integer $w\geq2$,
the formulas
%
%
\begin{equation}\label{Z-13}
\cases{
\displaystyle\overline{\mathcal U}_t=\sum_{p\geq1}c^2_{T_p} \bigl(
F''_{22}(\Delta X_{T_p},c_{T_p},c_{T_p})(V_p^-)^2\cr
\hspace*{59.2pt}{} + 2F''_{23}(\Delta X_{T_p},c_{T_p},c_{T_p})V_p^-V_p^+ \bigr)\cr
\displaystyle\hspace*{59.2pt}{} + F''_{33}(\Delta X_{T_p},c_{T_p},c_{T_p})(V_p^+)^2
1_{\{T_p\leq t\}},\cr
\displaystyle\overline{\mathcal U}{}'_t=\frac1{w^2} \sum_{p\geq1}c^2_{T_p} \bigl(
F''_{22}(\Delta X_{T_p},c_{T_p},c_{T_p})\bigl(V_p^-+\sqrt{w-1} V_p'^-\bigr)^2\cr
\hspace*{75.8pt}{} + \displaystyle2F''_{23}(\Delta X_{T_p},c_{T_p},c_{T_p})\cr
\hspace*{86.2pt}{}\times\displaystyle\bigl(V_p^-+\sqrt{w-1} V_p'^-\bigr)
\bigl(V_p^++\sqrt{w-1} V_p'^+\bigr) \bigr)\cr
\hspace*{75.8pt}{} + \displaystyle F''_{33}(\Delta X_{T_p},c_{T_p},c_{T_p})\bigl(V_p^++\sqrt{w-1} V_p'^+\bigr)^2
1_{\{T_p\leq t\}}}\hspace*{-38pt}
\end{equation}
define two c\`adl\`ag adapted processes $\overline{\mathcal U}$ and
$\overline{\mathcal U}{}'$ on the extended
filtered space $(\widetilde{\Omega},\widetilde{\mathcal
F},(\widetilde{\mathcal F}_t)_{t\geq0},\widetilde{\mathbb{P}})$.
Moreover, conditionally on $\mathcal F$, the pair
$(\overline{\mathcal U},\overline{\mathcal U}{}')$ is a process with
independent increments and finite
variation on compact intervals, and with the conditional means,
%
%
\begin{eqnarray}\label{Z-801}
\widetilde{\mathbb{E}}(\overline{\mathcal U}_t\mid\mathcal F) =
B'(F)_t,\qquad
\widetilde{\mathbb{E}}(\overline{\mathcal U}{}'_t\mid\mathcal F) =
\frac1w B'(F)_t
\nonumber\\[-8pt]\\[-8pt]
\eqntext{\mbox{where }\displaystyle B'(F)_t=\sum_{s\leq t}c_s^2 \bigl(F''_{22}(\Delta X_s,c_s,c_s)
+F''_{33}(\Delta X_s,c_s,c_s) \bigr).}
\end{eqnarray}
Here again, if we modify the exhausting sequence $(T_p)$ we accordingly
modify $\overline{\mathcal U}_t$ and $\overline{\mathcal U}{}'_t$, but
we \textit{do not change} their
$\mathcal F$-conditional laws.
\begin{theo}\label{TKP-3} Assume Assumption \ref{asssumHr}
for some $r<2$ and Assumption \ref{asssumKv} and
(\ref{Z5}) with
%
%
\begin{equation}\label{Z-119}
\rho<\bigl(2\varpi(2-r)\bigr)\wedge\frac{2v}{1+2v}\wedge\frac12.
\end{equation}
Let $F$ satisfy (\ref{Z-104}) for some $A\subset\mathbb{R}$, and
(\ref{Z-102}) for $\varepsilon=0$ when $r=0$ and some $\varepsilon>0$
otherwise.

\begin{longlist}
\item If either $r=0$, or $F(x,y,z)=0$ for $|x|\leq\varepsilon$ for
some $\varepsilon>0$,
the two-dimensional variables $ (k_nU(F,k_n)_t,k_nU(F,wk_n)_t )$
converge stably in law, in restriction to the set $\Omega^A_t$,
to the variable $(\overline{\mathcal U}_t,\overline{\mathcal U}{}'_t)$.

\item The same holds when $r>0$, provided $\rho$ and $\varpi$ satisfy
%
%
\begin{equation}\label{Z-113}
\rho< \bigl(\varpi(4-r)-1 \bigr)\wedge
\frac{(2v)\wedge1}{1+(2v)\wedge1}\wedge\frac12.
\end{equation}
\end{longlist}
\end{theo}

\eject
\section{\texorpdfstring{Construction of the tests.}{Construction of the tests}}\label{sec:TS}

\subsection{\texorpdfstring{Preliminaries.}{Preliminaries}}\label{ss:prel}
Now we are ready to construct our tests using the limit results of the
previous section. The overall interval on which the process $X$ is observed,
at times $i\Delta_n$, is $[0,T]$. In our tests the processes $X$
and $\sigma$ will not play a symmetrical role, mainly because $X$ is
observed, whereas $\sigma$ is not.

Although our main concern is to test for common jumps, irrespective of their
sizes, it might be useful to test also whether there are jumps of
$X$ with size in a subset $A$ of $\mathbb{R}$, occurring at the same
time as jumps
of $\sigma$: for example, $A=(a,\infty)$ or $A=(-\infty,-a)$
(positive or negative jumps of $X$ of size bigger than $a$ only),
or $A=(-\infty,-a)\cup(a,\infty)$ (jumps of $X$ of size bigger than $a$).

We thus pick a subset $A\subset\mathbb{R}$ satisfying the first part of
(\ref{Z-104}), and we are interested in the following two disjoint sets:
%
%
\begin{eqnarray}\label{Z201}
\qquad \Omega_{T}^{A,j}&=&\bigl\{\omega\dvtx\exists s\in(0,T] \mbox{ with }
\Delta X_s(\omega)\in A\setminus\{0\} \mbox{ and } \Delta\sigma
_s(\omega
)\neq0\bigr\},\nonumber\\
\Omega_{T}^{A,d}&=&\bigl\{\omega\dvtx\forall s\in(0,T],
\Delta X_s(\omega)\in A\setminus\{0\} \Rightarrow\Delta\sigma
_s(\omega
)=0,\\
&&\hspace*{51.8pt}\mbox{and } \exists s\in(0,T] \mbox{ with }
\Delta X_s(\omega)\in A\setminus\{0\}\bigr\}.\nonumber
\end{eqnarray}
The subscripts ``$j$'' and ``$d$'' stand for ``joint'' jumps and ``disjoint''
jumps. One could also specify a subset $A'$ in which the jumps of
$\sigma$
lie, but it requires more sophisticated CLTs than Theorems \ref{TKP-1}
and \ref{TKP-2}, and we will not consider this case here. Note that
$\Omega^{A,d}_T$ is contained in the set $\Omega^A_T$ of Theorem \ref{TKP-3}.

Next, we recall that testing a null hypothesis ``we are in
a subset $\Omega_{0} $'' of $\Omega$, against the
alternative ``we
are in a subset $\Omega_{1}$,'' with of course $\Omega_{0}
\cap\Omega_{1}=\varnothing$, amounts to finding a critical
(rejection) region
$C_{n}\subset\Omega$ at stage $n$. The asymptotic size and asymptotic
power for this sequence $(C_n)$ of critical regions are the following
numbers:
%
%
\begin{eqnarray}\label{Z-14}
\alpha &=& \sup\Bigl( \limsup_{n} \mathbb{P}(C_{n}\mid H)\dvtx H\in\mathcal F,
H\subset\Omega_{0}, \mathbb{P}(H)>0 \Bigr),\nonumber\\[-8pt]\\[-8pt]
\beta &=& \inf\Bigl( \liminf_{n} \mathbb{P}(C_{n}\mid H)\dvtx H\in\mathcal F,
H\subset\Omega_{1}, \mathbb{P}(H)>0 \Bigr).\nonumber
\end{eqnarray}

In all forthcoming tests, we fix a priori two sequences
$u_n$ and $k_n$ satisfying (\ref{Z5}): typically
$u_n=a\Delta_n^\varpi$ and $k_n=[a'/\Delta_n^\rho]$ where $a,a'>0$ are
constants. Some restrictions on $\varpi$ and $\rho$ will also be made,
depending on the test at hand.

Finally, similar to the tests for deciding whether price and
volatility jump
together or not which we develop here, one can use the limit results of
Section~\ref{sec:KP} to derive various other tests about the relationship
between jumps in $X$ and its volatility. Examples include: (1) testing
whether all jumps in $X$ are associated with volatility jumps and (2)
testing whether jumps in $X$ of given sign always lead to positive (negative)
volatility jumps.

\subsection{\texorpdfstring{Testing the null hypothesis ``no common jump.''}{Testing
the null hypothesis ``no common jump''}}

Here we take the null hypothesis to be ``$X$ and $\sigma$ have no
common jump'' with jump size of $X$ in $A$, that is, $\Omega
_T^{(A,d)}$, for
$A$ like in (\ref{Z-104}).

\subsubsection{\texorpdfstring{General family of tests.}{General family of tests}}
The idea is to use the variable $U(F)_T$ of (\ref{Z0}) and its
approximations $U(F,k_n)_T$ for a suitable function $F$, namely
%
%
\begin{eqnarray}\label{TE-1}
F(x,y,z) = f(x)g(y,z) \nonumber\\
\eqntext{\mbox{with }\cases{
f \mbox{ is $C^1$},\qquad \mbox{on } R,\cr
\mbox{$x\in[-\varepsilon,\varepsilon] \quad\Rightarrow\quad|f'(x)|\leq C|x|^{p-1}$},\cr
x\in A\setminus\{0\} \quad\Rightarrow\quad f(x)>0,\cr
\mbox{$x\notin A\setminus\{0\} \quad\Rightarrow\quad f(x)=0$,}}}
\\[-37pt]
\\[20pt]
\eqntext{\cases{
g \mbox{ is $C^2$},\qquad\mbox{with bounded first}\cr
\hspace*{4.6pt}\phantom{g \mbox{ is $C^2$}}\qquad\mbox{and second derivatives},\cr
z\neq y \quad\Rightarrow\quad g(y,z)>0,\cr
z=y \quad\Rightarrow\quad g(y,z)=0,\cr
g'_{1}(y,y)=g'_{2}(y,y)=0,\cr
g''_{11}(y,y)+g''_{22}(y,y)>0,}}
\end{eqnarray}
and where $p\geq1\vee r$. These ensure that $F$ satisfies (\ref{Z-101}),
(\ref{Z-102}) and (\ref{Z-104}). It also implicitly implies\vspace*{1pt} conditions
on the set $A$, since $A\setminus\{0\}=f^{-1}((0,\infty))$
and $f$ is $C^1$ on $R$, whose complement is finite.

By Theorem \ref{TKP-1}, we have the following convergence:
%
%
\begin{equation}\label{TE-3}
U(F,k_n)_T \stackrel{\mathbb{P}}{\longrightarrow}U(F)_T
\cases{
=0, &\quad  on the set $\Omega_T^{(A,d)}$,\vspace*{2pt}\cr
>0, &\quad on the set $\Omega_T^{(A,j)}$.}
\end{equation}
So in order to test the null hypothesis $\Omega_T^{(A,d)}$,
it is natural at stage $n$ to take a critical region of the form
$C_n=\{U(F,k_n)_T>Z_n\}$ for some (possibly random) $Z_n>0$.
In order to determine $Z_n$ in such a way that the asymptotic level of
the test be some $\alpha$, we make use of Theorem \ref{TKP-3}, which says
that, in restriction to the set $\Omega_T^{(A,d)}$, the variables
$k_nU(F,k_n)_T$ converge stably in law to $\overline{\mathcal U}_T$,
as defined by
(\ref{Z-13}). Conditionally on $\mathcal F$, this variable is a weighted
chi-square variable, with mean $B'(F)_T$ given by (\ref{Z-801}).

One simple, not very efficient, way to derive test with a
prescribed level $\alpha$ makes use of Bienaym\'e--Chebyshev inequality,
plus the fact that by Theorem \ref{TKP-1} again we can approximate
the variable $B'(F)$ by $U(G,k_n)_T$ where
%
%
\begin{equation}\label{TE-4}
G(x,y,z) = y^2 f(x) \bigl(g''_{11}(y,z)+g''_{22}(y,z)\bigr)
\end{equation}
satisfies all the requirements of that theorem. At this point, the critical
region is taken to be
%
%
\begin{equation}\label{TE-5}
C_n = \biggl\{U(F,k_n)_T > \frac{U(G,k_n)_T}{\alpha k_n} \biggr\}
\end{equation}
and the following is straightforward:
\begin{theo}\label{TTE-1} Assume Assumptions \ref{asssumHr} and \ref{asssumKv}, and $F$ as in
(\ref{TE-1}) with $p\geq r$, and
choose $u_n$ and $k_n$ such that (\ref{Z5}) and (\ref{Z-113}) hold.
Then the critical region (\ref{TE-5}) has asymptotic level less than
$\alpha$
for testing the null hypothesis $\Omega_T^{(A,d)}$, and asymptotic
power $1$ for the alternative $\Omega_T^{(A,j)}$.
\end{theo}

The actual asymptotic size of this test is usually much lower than
$\alpha$,
because Bienaym\'e--Chebyshev is a crude approximation. However we can
use a Monte Carlo simulation to better fit the size, in the spirit of
\cite{JT}: we take a sequence $N_n\to\infty$, and we
simulate independent $\mathcal N(0,1)$ variables $V^-_i(j)$ and
$V^+_i(j)$ of independent $\mathcal N(0,1)$ variables, for $j=1,\ldots
, N_n$
and $i=1,\ldots,[T/\Delta_n]$. Then, with the observed values of
$\Delta^n_iX$,
hence of the variables $\widehat{c}(k_n)_i$ as well, we set
%
%
\begin{eqnarray}\label{TE-20}\quad
\overline{\mathcal U}(n,j)&=&\sum_{i=k_n+1}^{[T/\Delta_n]-k_n}
f(\Delta^n_iX)1_{\{|\Delta^n_iX|>u_n\}} (\widehat{c}(k_n)_i)^2\nonumber\\
&&\hspace*{40.72pt}{}\times \bigl(g''_{11}\bigl(\widehat{c}(k_n)_{i-k_n-1},\widehat
{c}(k_n)_i\bigr)(V^-_i(j))^2\nonumber\\[-8pt]\\[-8pt]
&&\hspace*{40.72pt}\hspace*{14.98pt}{} +g''_{22}\bigl(\widehat{c}(k_n)_{i-k_n-1},\widehat
{c}(k_n)_i\bigr)(V^+_i(j))^2\nonumber\\
&&\hspace*{40.72pt}\hspace*{14.98pt}{} +2g''_{12}\bigl(\widehat{c}(k_n)_{i-k_n-1},\widehat
{c}(k_n)_i\bigr)V^-_i(j)V^+_i(j) \bigr).\nonumber
\end{eqnarray}
Next, we consider the order statistics of these simulated variables,
that is, $\overline{\mathcal U}(n)_{(1)}\geq\overline{\mathcal
U}(n)_{(2)}\geq\cdots\geq\overline{\mathcal U}(n){(N_n)}$
such that $\{\overline{\mathcal U}(n)_j\dvtx1\leq j\leq N_n\}=\{\overline
{\mathcal U}(n,j)\dvtx 1\leq j\leq N_n\}$,
and we take as our critical region the following:
%
%
\begin{equation}\label{TE-6}
C_n = \biggl\{U(F,k_n)_T > \frac{\overline{\mathcal U}(n)_{([N_n\alpha
])}}{k_n} \biggr\}.
\end{equation}
\begin{theo}\label{TTE-2} Assume Assumptions \ref{asssumHr} and \ref{asssumKv}, and $F$ as in
(\ref{TE-1}) with $p\geq r$, and
choose $u_n$ and $k_n$ such that (\ref{Z5}) and (\ref{Z-113}) hold.
Then the critical region (\ref{TE-6}), constructed with any sequence
$N_n$ increasing to infinity, has asymptotic level equal to $\alpha$ for
testing the null hypothesis $\Omega_T^{(A,d)}$, and asymptotic
power $1$ for the alternative $\Omega_T^{(A,j)}$.
\end{theo}

\subsubsection{\texorpdfstring{A leading example.}{A leading example}}
Here we specialize $A$ to be either $A=\mathbb{R}$ or $A=[-a,a]^c$ for
some positive
$a$, and in the first case we will need $r=0$; that
is, our process $X$ has finite activity jumps. In both cases, we end up
using a
finite number of jumps of $X$ (jumps of size higher than a fixed value are
almost surely of finite number); therefore we consider $F(x,y,z)=f(x)g(y,z)$
with $f(x)=1_{\{x\in A\}}$. Since for this choice $f(x)$ is
discontinuous at
$x=\pm a$, we need $\pm a\notin D$ [recall (\ref{Z701})] in order for
(\ref{Z-102}) to be satisfied. Of course, $D$ is unknown, but in the
typical case when the L\'evy measure of $X$ has no atom, $D=\{0\}$ and
thus any $a>0$ works. Otherwise, we can replace $1_{\{|x|>a\}}$ by a $C^1$
function which is very close to this. Practically this should make no
significant difference, and therefore we stick to the indicator function,
with $a\notin D$. When $A=\mathbb{R}$ we set $a=0$.

A natural choice for the function $g$ is the following:
%
%
\begin{equation}\label{TE-7}
g(y,z) = 2\log\frac{y+z}2-\log y-\log z.
\end{equation}
This choice corresponds to the log-likelihood ratio test for testing
that two
independent samples of i.i.d. zero-mean normal variables have the same
variance. The link with our testing comes from the fact that around a jump
time the high-frequency increments of $X$ are ``approximately'' i.i.d.
normal.

With this choice of $F$, our test for common jumps becomes essentially
pivotal, that is, the limiting distribution of the test statistics
depends only
on the number of jumps and is thus straightforward to implement. To see
this, note that in this case (\ref{Z-13}) writes as
%
%
\begin{equation}\label{TE-8}
\overline{\mathcal U}_T = \frac12\sum_{p\geq1}(V^+_p-V^-_p)^2 1_{\{
|\Delta X_{T_p}|>a\}}.
\end{equation}
Conditionally on $\mathcal F$, this variable has the same law as a chi-square
variable with $N_T$ degrees of freedom where
$N_T=\sum_{p\geq1}1_{\{|\Delta X_{T_p}|>a\}}$.
The variable $N_T$ is not observable. However, we have
%
%
\begin{equation}\label{TE-9}
N_T^n = \sum_{i=1}^{[T/\Delta_n]} 1_{\{|\Delta^n_iX|>a\vee u_n\}}
\stackrel{\mathbb{P}}{\longrightarrow}N_T,
\end{equation}
and since these are integer-valued variables we even have $\mathbb{P}
(N^n_T=N_T)\to1$.

Therefore, denoting by $z(\alpha,n)$ the $\alpha$-quantile of a
chi-square variable
$\chi^2_n$ with $n$ degrees of freedom, that is, the number such that
$\mathbb{P}(\chi^2_n>z(\alpha,n))=\alpha$, we may take the
following critical
region at
stage $n$:
%
%
\begin{equation}\label{TE-10}
C_n = \biggl\{U(F,k_n)_T>\frac{z(\alpha,N^n_t)}{k_n} \biggr\}.
\end{equation}
\begin{theo}\label{TTE-3} Assume Assumptions \ref{asssumHr} and \ref{asssumKv}, and $F$ as above
with either $a=0$ if $r=0$ or $a$ positive and $\pm a\notin D$ if
$r\geq0$.
Choose $u_n$ and $k_n$ such that (\ref{Z5}) and (\ref{Z-119}) hold. Then
the critical region (\ref{TE-9}) has asymptotic level equal to $\alpha
$ for
testing the null hypothesis $\Omega_T^{(A,d)}$, and asymptotic power
$1$ for
the alternative~$\Omega_T^{(A,j)}$.
\end{theo}

Note that for constructing the critical region in (\ref{TE-10}), we
need only
the critical values of a chi-square variable $\chi^2_n$, and thus
there is no
need for simulation.

\subsection{\texorpdfstring{Testing the null hypothesis, ``common
jump.''}{Testing the null hypothesis, ``common jump''}}

Now we take the null hypothesis to be ``$X$ and $\sigma$ have common
jumps'' with sizes in $A$ for $X$, that is, $\Omega_T^{(A,j)}$, for
$A$ like in (\ref{Z-104}). We take an integer $w\geq2$ and a
function $F$ satisfying (\ref{TE-1}), and introduce the statistics
%
%
\begin{equation}\label{TE-11}
S_n = \frac{U(F,wk_n)_T}{U(F,k_n)_T}.
\end{equation}

If we combine Theorems \ref{TKP-1} and \ref{TKP-3}, we first obtain
%
%
\begin{equation}\label{TE-12}
\cases{
S_n \stackrel{\mathbb{P}}{\longrightarrow}1, &\quad on the set
$\Omega_T^{(A,j)}$,\cr
S_n \stackrel{\mathcal L-(s)}{\longrightarrow} \dfrac{\overline
{\mathcal U}{}'_T}{\overline{\mathcal U}_T} \neq1 \mbox{ a.s.},
&\quad on the set $\Omega_T^{(A,d)}$,}
\end{equation}
where $ \stackrel{\mathcal L-(s)}{\longrightarrow} $ stands for the
stable convergence in law; for
the second
convergence we must assume that $k_n$ satisfies (\ref{Z-119}), and
$\overline{\mathcal U}{}'_T$
is implicitly depending on $w$; note that the
pair $(\overline{\mathcal U}_T,\overline{\mathcal U}{}'_T)$ has
$\mathcal F$-conditionally a density, implying
$\overline{\mathcal U}{}'_T/\overline{\mathcal U}_T\neq1$ a.s.

To determine the asymptotic level of a test based upon $S_n$, we make
use of Theorem \ref{TKP-2} which by way of the delta method shows that,
in restriction to the set $\Omega_T^{(A,j)}$, the variables
$\sqrt{k_n} (S_n-1)$ converge stably in law to $ (\sqrt{w-1}
\mathcal U'_T-
(w-1)\mathcal U_T )/wU(F)_T$. The limit is $\mathcal F$-conditionally centered
Gaussian with variance $(w-1)B(F)_T/w(U(F)_T)^2$ [recall (\ref{Z-8})].
Hence, if
%
%
\begin{eqnarray}\label{TE-13}
G(x,y,z) &=& 2f(x)^2 \bigl(y^2 g'_1(y,z)^2+z^2 g'_2(y,z)^2
\bigr),\nonumber\\[-8pt]\\[-8pt]
V_n &=& \frac{(w-1) U(G,k_n)_T}{w k_n (U(F,k_n)_T)^2},\nonumber
\end{eqnarray}
we deduce that, in restriction to the set $\Omega_T^{(A,j)}$, the variables
$(S_n-1)/\sqrt{V_n}$ converge stably in law to a standard normal
variable, under (\ref{Z-103}), of course.

Then we may take the following critical region at stage $n$, where
$z_\alpha$ denotes the symmetric $\alpha$-quantile of an
$\mathcal N(0,1)$ variable $V$, that is, $\mathbb{P}(|V|>z_\alpha
)=\alpha$.
%
%
\begin{equation}\label{TE-14}
C_n = \bigl\{|S_n-1|>z_\alpha\sqrt{V_n} \bigr\}.
\end{equation}
\begin{theo}\label{TTE-4} Assume Assumptions \ref{asssumHr} and \ref{asssumKv}, and $F$ as in
(\ref{TE-1}) with $p>1+r/2$. Choose $u_n$
and $k_n$ such that (\ref{Z5}) and (\ref{Z-103}) hold. Then
the critical region (\ref{TE-14}) has asymptotic level $\alpha$ for
testing the null hypothesis $\Omega_T^{(A,j)}$.
\end{theo}

There is no statement about the asymptotic power for the alternative
$\Omega_T^{(A,d)}$ which is any case is \textit{not} equal to $1$.
Indeed, on
$\Omega_T^{(A,d)}$, the variables $(S_n-1)/\sqrt{V_n}$ converge stably
in law to
some limit $\mathcal V$ (easily constructed from $\overline{\mathcal
U}_T$, $\overline{\mathcal U}{}'_T$ and also
the variable $\overline{\mathcal U}_T$ associated with the function
$G$) as soon as $G$
satisfies the assumption of Theorem \ref{TKP-3}. The variable
$\mathcal V$
is a.s.
nonvanishing, and the asymptotic power of our test is
\[
\beta= \inf\bigl(\mathbb{P}(|\mathcal V|>z_\alpha\mid H)\dvtx H\in\mathcal
F, H\subset
\Omega_T^{(A,d)}, \mathbb{P}(H)>0\bigr).
\]
This quantity cannot be computed explicitly and may be close to $0$,
as simulations show later on.

To avoid this power problem, we can ``truncate'' the estimated
variance $V_n$: let $v_n$ be a sequence of positive numbers (possibly
random, but of course depending only on the observations at stage $n$),
such that $v_n\to0$ and $k_nv_n\to\infty$, and set
\[
V'_n = V_n\wedge v_n.
\]
Since $k_nV_n$ converges to a positive finite limit on $\Omega_T^{(A,j)}$,
we have $\mathbb{P}(V_n=V'_n)\to1$ and this truncation has no effect
on the
behavior of our standardized statistics under the null,
and we take the following critical region:
%
%
\begin{equation}\label{TE-15}
C'_n = \bigl\{|S_n-1|>z_\alpha\sqrt{V'_n} \bigr\}.
\end{equation}
\begin{theo}\label{TTE-5} Assume Assumptions \ref{asssumHr} and \ref{asssumKv}, and $F$ as in
(\ref{TE-1}) with $p>1+r/2$. Choose $u_n$
and $k_n$ such that (\ref{Z5}) and (\ref{Z-119}) hold. Then if
$v_n\to0$ and $k_nv_n\to\infty$,
the critical region (\ref{TE-15}) has asymptotic level $\alpha$ for
testing the null hypothesis $\Omega_T^{(A,j)}$, and asymptotic power $1$
for the alternative $\Omega_T^{(A,d)}$.
\end{theo}
\begin{rem}\label{RTE-1} Exactly as in the previous subsection,
when $r=0$ we may use the function $F(x,y,z)=g(y,z)$ given by
(\ref{TE-7}), and $A=\mathbb{R}$. When $r>0$ we can use $F(x,y,z)=g(y,z)
1_{\{|x|>a\}}$, with $g$ as above and $a>0$ and $A=[-a,a]^c$, provided
$\pm a\notin D$. In these cases, $\rho$ and $\varpi$
are subject to the weaker condition (\ref{Z-129}) only.
\end{rem}

\subsection{\texorpdfstring{Practical aspects.}{Practical aspects}}
The construction of the tests involves several choi\-ces to be made by the
user.
The first one is about the functions $f$ and $g$ in (\ref{TE-1}). A~good
choice seems to be $f(x)=1_{\{|x|>a\}}$ for some $a\geq0$ and $g$ as given
by (\ref{TE-7}). However this works only when (H-$0$) holds (a serious
restriction indeed), or when $a>0$, and in the latter case we only test
for common jumps when the size of the jumps of $X$ is bigger than $a$.
Then the user can perform the testing for various levels of $a$. In addition,
if jumps of certain size in $X$ are more important, $1_{\{|x|>a\}}$ can be
replaced with an appropriate weighting function for the jumps of different
size. Finally, if the user wants to check cojumping, including the very
``small'' jumps in~$X$, then a good choice is to take $f(x)=x^2$ and
$g(y,z)=h(y-z)$ where $h$ is a $C^2$ function with bounded first and second
derivatives, and $h(0)=h'(0)=0$ and $h''(0)>0$ and $h(x)>0$ when $x\neq0$.

The second choice in implementing the tests is about the sequences
$u_n$ and~$k_n$. Here we face a natural tradeoff between efficiency and robustness.
$u_n$ and $k_n$ should satisfy (\ref{Z-129}) or (\ref{Z-119}) when
$f(x)=1_{\{|x|>a\}}$, and (\ref{Z-103}) or (\ref{Z-113}) otherwise,
depending on which test is performed. These conditions depend on the a
priori unknown numbers $r$ and $v$ in Assumptions \ref{asssumHr} and \ref{asssumKv}. The higher
the $r$ and the lower the $v$ are, the stricter the conditions are, and
the lower the rate at which $k_n$ can grow, that is, the slower the
rate at
which $U(F,k_n)_T$ converges. Intuitively, high $r$ makes it difficult to
distinguish the many small jumps from the Brownian increments, while low
$v$ means volatility is very ``active'' over short intervals and that makes
estimation from neighboring increments ``noisier.''

Most stochastic volatility models imply that $\sigma_t$ is an It\^o
semimartingale and therefore $v=\frac12$. If in addition we assume that
$r<1$, that is, jumps are of finite variation, then we can choose
$\varpi$ and
$\rho$ arbitrarily close to $\frac12$, which is the optimal choice.
Alternatively, if we are willing to assume only that $r\leq r_0$ for some
$1<r_0<2$, then we can write the conditions on $\varpi$ and $\rho$ with
respect to $r_0$ and pick $u_n$ and $k_n$ so that they are fulfilled. One
should emphasize that $\varpi$ and $\rho$ only give an order of magnitude,
and the concrete choice of $u_n$ and $k_n$ when one is faced with a set of
data and thus with $n$ and $\Delta_n$ given is always a difficult
question: in
the Monte Carlo study we provide some guidance on that.

The last choice to be made, for the second test, is choosing the
integer $w$.
Under the null $\Omega_T^{(A,j)}$ the normalized asymptotic $\mathcal
F$-conditional
variance of $S_n$ takes the form $\frac{w-1}{w} \Phi$ where $\Phi=
B(F)_Y/(U(F)_T)^2$ does not depend on $w$. The
minimum of $\frac{w-1}{w}$ for $w\geq2$ is achieved at $w=2$.
At the same time the effect of changing $w$ under the alternative
hypothesis is unclear and in general depends on the particular
realization. For that reason we suggest to take $w=2$ and we do so in
our numerical applications without further mention.
Some Monte Carlo experiments (not reported here) with $w=4$ provide
further support for this choice.

\section{\texorpdfstring{Monte Carlo study.}{Monte Carlo study}}\label{sec:MC}

In this section we check the performance of our tests on simulated data.
We work with the stochastic volatility model
%
%
\begin{eqnarray}
dX_{t}&=& \sqrt{V_t^1+V^2_t}\,dW_{t}+\alpha_0\int_{\mathbb{R}}x\mu(dt,dx,dy),
\nonumber\\
dV^1_t&=&\kappa_1(\theta-V_t^1)\,dt+\sigma\sqrt{V^1_t}\,dW_t',\\
dV_t^2&=&-\kappa_2V_t^2\,dt+\alpha_1\int_{\mathbb{R}}y\mu(dt,dx,dy)
+\alpha_2\int_{\mathbb{R}}y\mu'(dt,dy),\nonumber
\end{eqnarray}
where $W$ and $W'$ are two independent Brownian motions; the (finite activity)
Poisson measures $\mu$ and $\mu'$ are independent with
compensators\break
$\nu(dt,dx,dy)=\frac{\lambda}{2(h-d)(u-d)} 1_{(x\in[-h;-l]\cup
[l;h])}
1_{(y\in[d;u])} \,dt\,dx\,dy$ for $0<l<h$ and $0<d<u$ and
$\nu'(dt,dy)=\frac{\lambda}{u-d} 1_{(y\in[d;u])} \,dt\,dy$. This two-factor
volatility structure is found to fit high-frequency financial data very
well in \cite{T09} (see also references therein). The above cited study
finds the continuous volatility factor to be very persistent, while the
discontinuous one to be transient. This is reflected in our choice of the
parameter values of $\kappa_1$ and $\kappa_2$ in the Monte Carlo settings,
in an effort to make them realistically plausible for financial applications.
In Table \ref{tb:ps} we report the parameter values for all cases considered.
In all of them the variance of the jumps in $X$ is fixed and its share in
the total price variation is in the range $0.2-0.34$, which is similar to
one estimated from real financial data (see, e.g., \cite{HT}).
Scenarios with
a higher number of jumps imply that the jumps are of smaller size. The different
parameter settings differ in the average number of jumps, their sizes,
whether jumps are present in the volatility and whether they arrive
together with the jumps in $X$ or not. The cases labeled with $c$ and $d$
are draws from the set $\Omega_T^{(A,d)}$, while the cases labeled
with $j$
and $m$ are draws from the set $\Omega_T^{(A,j)}$. To ensure the
latter, we
discard simulations from scenarios $m$ on which there is no common
price and
volatility jumps. The behavior of the tests on the discarded simulation draws
is exactly as on the simulations from scenarios $d$.

%
\begin{table}
\caption{Parameter settings used in the Monte Carlo}
\label{tb:ps}
\begin{tabular*}{\tablewidth}{@{\extracolsep{\fill}}lcccccccccccc@{}}
\hline
&\multicolumn{12}{c@{}}{\textbf{Parameters}}\\[-4pt]
&\multicolumn{12}{c@{}}{\hrulefill} \\
\textbf{Case} & $\bolds{\kappa_1}$ & $\bolds\theta$ & $\bolds\sigma$
& $\bolds{\kappa_2}$ &$\bolds{\alpha_0}$ &
$\bolds{\alpha_1}$& $\bolds{\alpha_2}$ & $\bolds\lambda$
& $\bolds l$ & $\bolds h$ & $\bolds d$ &
$\bolds u$\\
\hline
I-c & 0.02& 0.4& 0.04& 0.5& 1& 0& 0& 0.5& 0.1& 1.0420& & \\
II-c & 0.02& 0.4& 0.04& 0.5& 1& 0& 0& 1.0& 0.1& 0.7197& & \\
III-c& 0.02& 0.4& 0.04& 0.5& 1& 0& 0& 4.0& 0.1& 0.3275& & \\
I-d & 0.02& 0.4& 0.04& 0.5& 1& 0& 1& 0.5& 0.1& 1.0420& 0.04 & 0.7600\\
II-d & 0.02& 0.4& 0.04& 0.5& 1& 0& 1& 1.0& 0.1& 0.7197& 0.04 & 0.3600\\
III-d& 0.02& 0.4& 0.04& 0.5& 1& 0& 1& 4.0& 0.1& 0.3275& 0.04 & 0.0600\\
I-j & 0.02& 0.4& 0.04& 0.5& 1& 1& 0& 0.5& 0.1& 1.0420& 0.04 & 0.7600\\
II-j & 0.02& 0.4& 0.04& 0.5& 1& 1& 0& 1.0& 0.1& 0.7197& 0.04 & 0.3600\\
III-j& 0.02& 0.4& 0.04& 0.5& 1& 1& 0& 4.0& 0.1& 0.3275& 0.04 & 0.0600\\
I-m & 0.00& 0.0& 0.00& 0.5& 1& 1& 1& 0.5& 0.1& 1.0420& 0.04 & 0.7600\\
II-m & 0.00& 0.0& 0.00& 0.5& 1& 1& 1& 1.0& 0.1& 0.7197& 0.04 & 0.3600\\
III-m& 0.00& 0.0& 0.00& 0.5& 1& 1& 1& 4.0& 0.1& 0.3275& 0.04 & 0.0600\\
\hline
\end{tabular*}
\end{table}

In the simulated model we have (H-$0$) and (K-$1/2$), so we use the tests
based on $f(x)=1$ and $g$ given by (\ref{TE-7}), and $A=\mathbb{R}$.
Throughout, time is measured in days, and the
observation length is five days, that is, $T=5$, which constitutes one
business week. We simulate $5000$ days, that is, $1000$ Monte Carlo
replications.
On each day we consider sampling $n=1000$, $n=5000$ or $n=24\mbox{,}000$
times, corresponding approximately to sampling every $0.5$
minutes, $5$ seconds or $1$ second for a trading day of $6.5$
hours or equivalently to sampling every $1.5$ minutes, $15$
seconds or $4$ seconds for a trading day of $24$ hours. Finally,
for the calculation of the local volatility estimators we use a
window $k_n=[5\times\Delta_n^{-0.49}]$. Our choice for the
truncation parameters $a$ and $\varpi$ determining $u_n=a\Delta
_n^\varpi$ is
$a=5\times\sqrt{\mathrm{BV}}$ and $\varpi=0.49$, respectively, where $\mathrm{BV}$
denotes the bi-power variation over the day \cite{BS1,BS4}. This
choice of
the truncation level reflects the time-variation in the volatility.

%
\begin{figure}[b]

\includegraphics{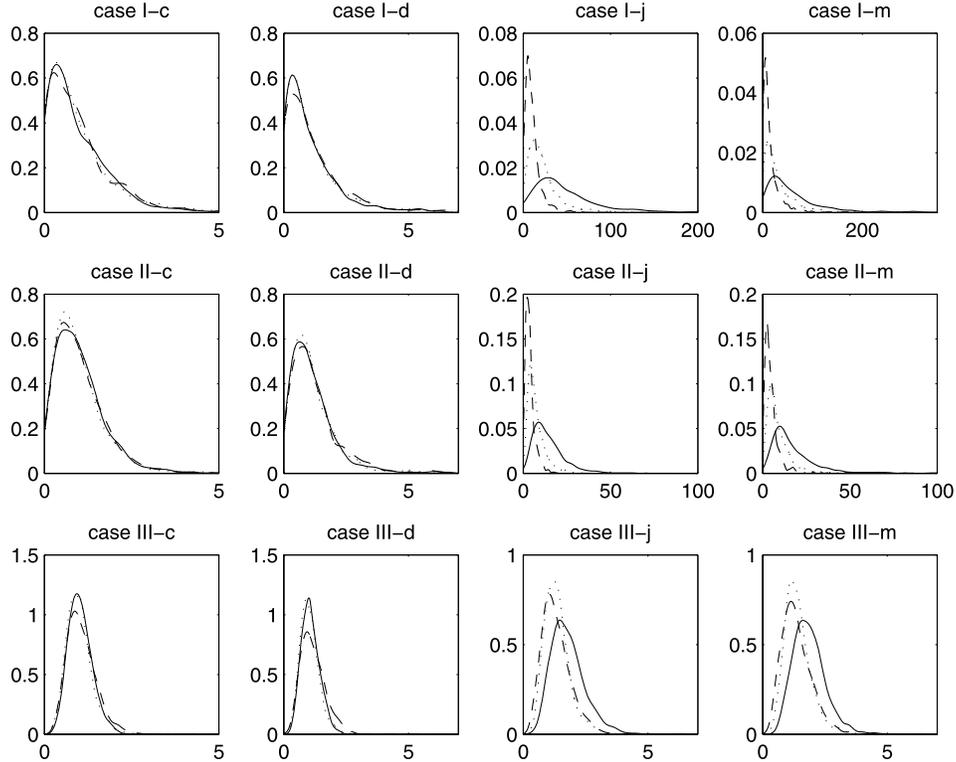}

\caption{Kernel density estimate of $U(f,g,k_n)_T/N_t^n$ from the
Monte Carlo. The dashed line corresponds to sampling frequency of
$n=1000$, the dotted line to sampling frequency of $n=5000$ and
the solid line to sampling frequency of $n=24\mbox{,}000$.}
\label{fig:hist_d}
\end{figure}

%
\begin{figure}

\includegraphics{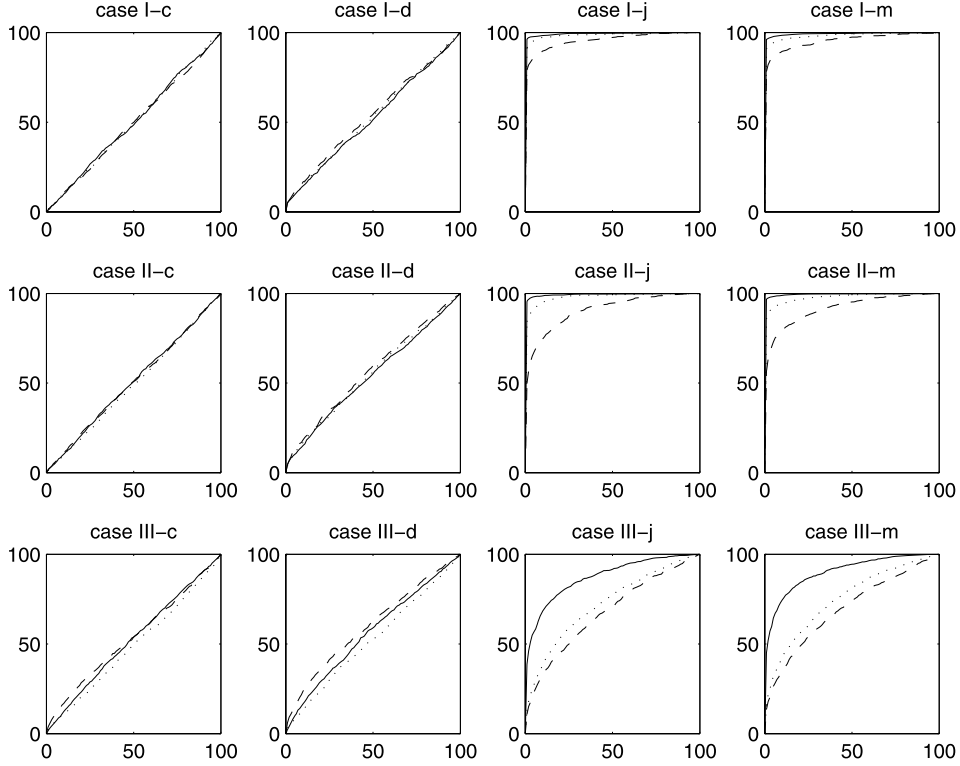}

\caption{Size and power of the test for disjoint price and
volatility jumps. The x-axis shows the nominal level of the
corresponding test, while the y-axis shows the percentage of rejection
in the Monte Carlo. The dashed line corresponds to sampling frequency
of $n=1000$, the dotted line to sampling frequency of $n=5000$ and the
solid line to sampling frequency of $n=24\mbox{,}000$.} \label{fig:test_d}
\end{figure}

%
\begin{figure}

\includegraphics{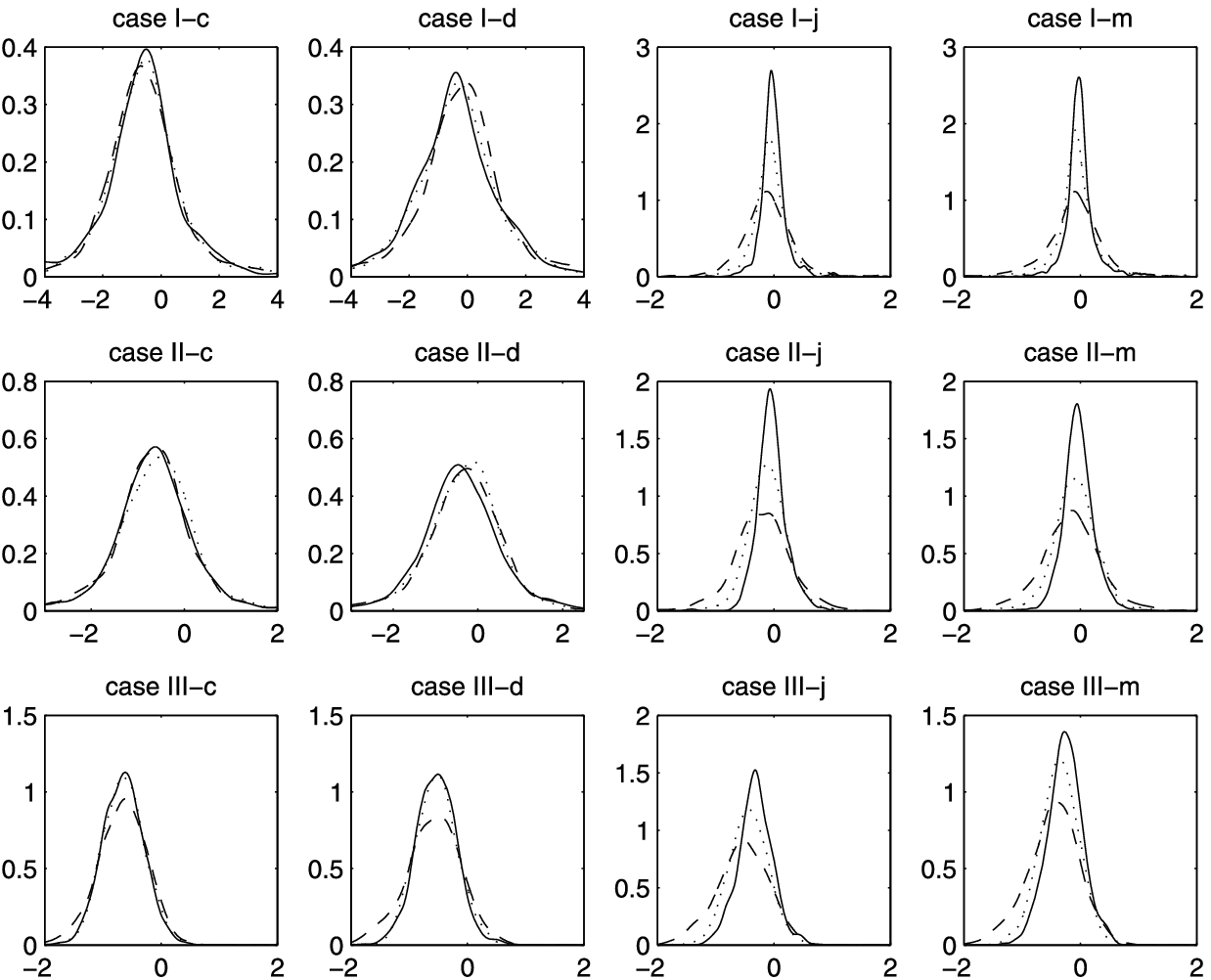}

\caption{Kernel density estimate of $\log(S_n)$ from the Monte Carlo.
The dashed line corresponds to sampling frequency of
$n=1000$, the dotted line to sampling frequency of $n=5000$ and
the solid line to sampling frequency of $n=24\mbox{,}000$.}
\label{fig:hist_j}
\end{figure}

%
\begin{figure}

\includegraphics{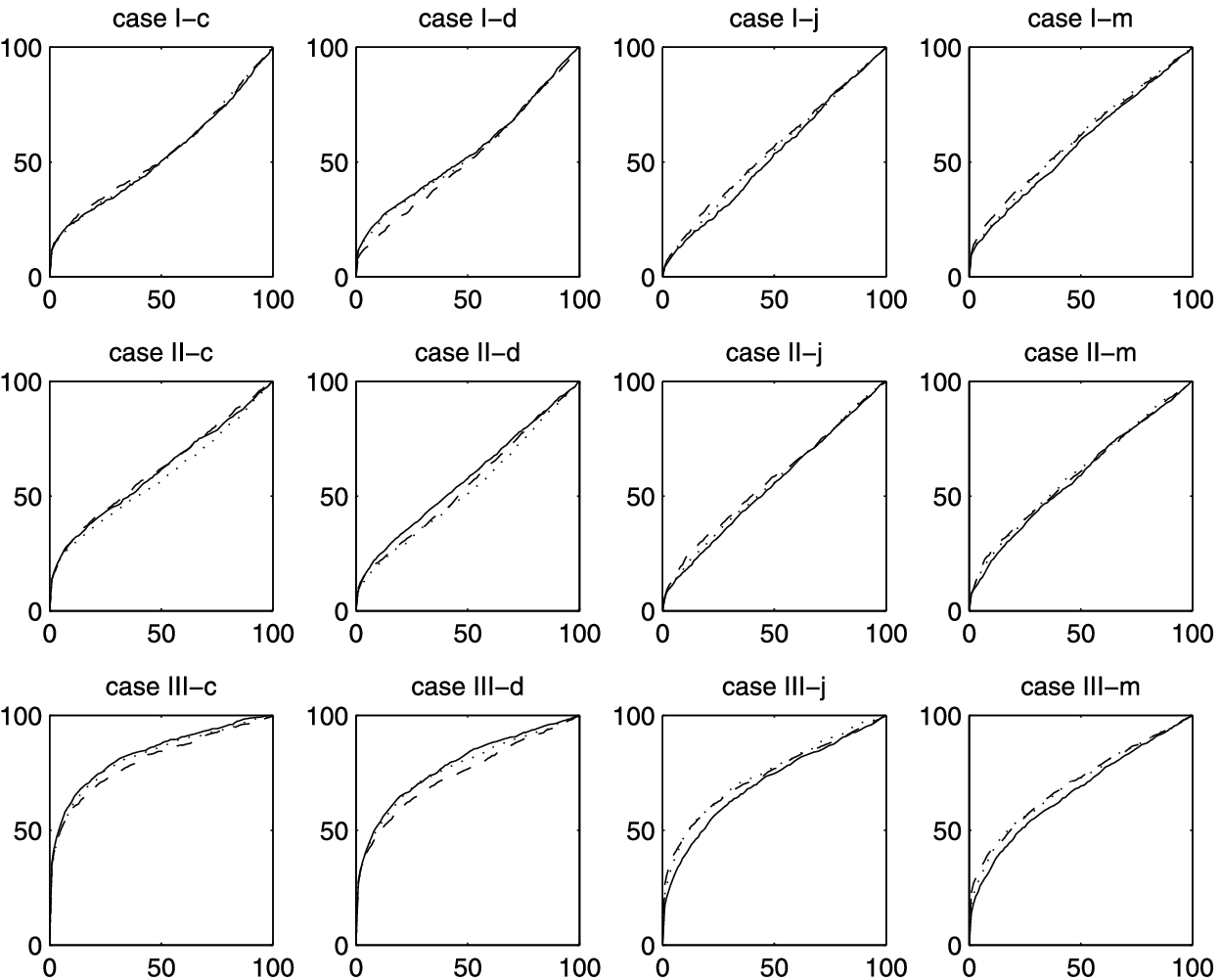}

\caption{Size and power
of the test for common price and volatility jumps with $V_n$ used in the
construction of the critical region. The x-axis shows the nominal level of
the corresponding test, while the y-axis shows the percentage of rejection
in the Monte Carlo.
The dashed line corresponds to sampling frequency of $n=1000$, the dotted
line to sampling frequency of $n=5000$ and the solid line to sampling
frequency of $n=24\mbox{,}000$.} \label{fig:test_j}
\end{figure}

%
\begin{figure}

\includegraphics{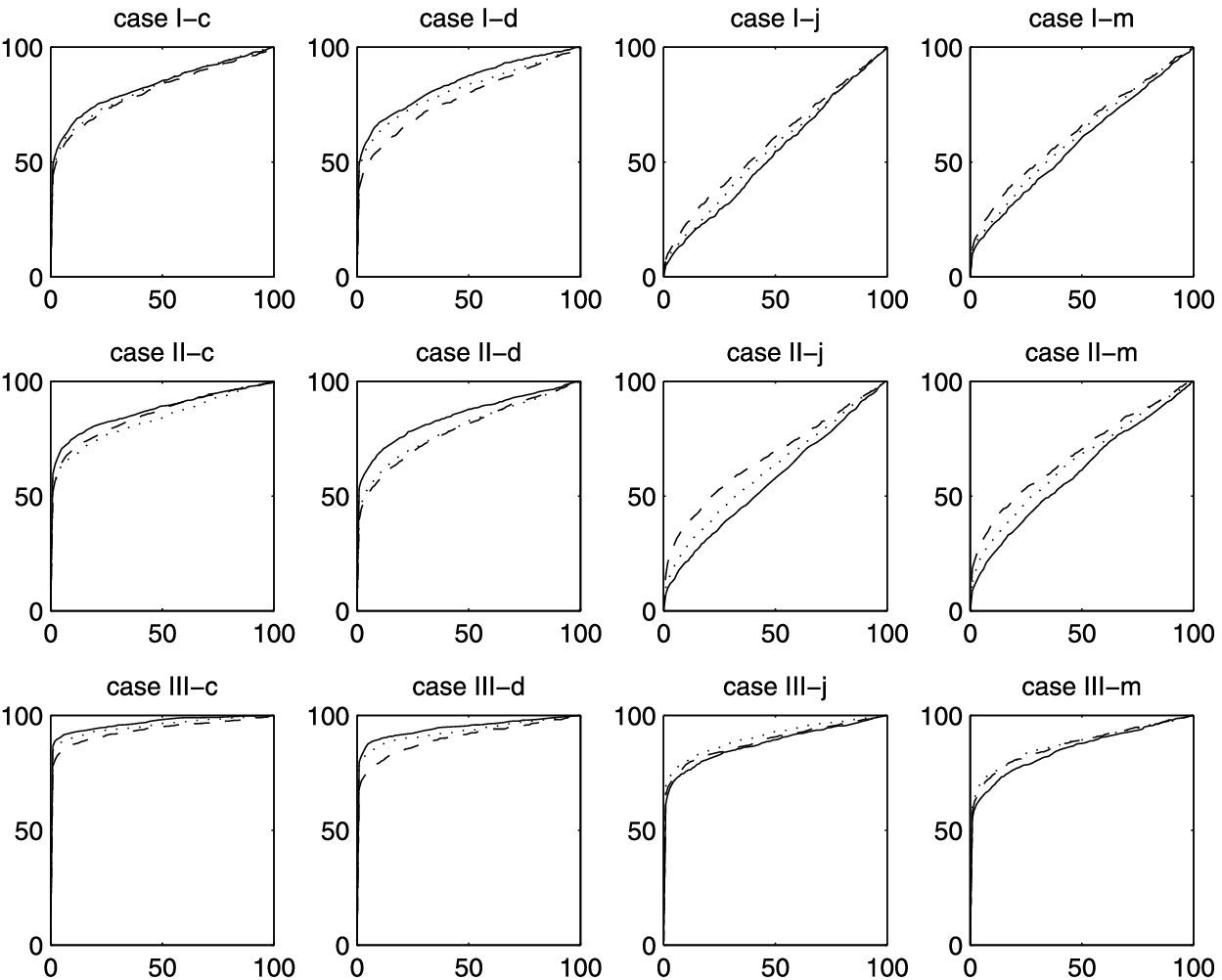}

\caption{Size and power of the test for common price and volatility jumps
with $V'_n$ used in the construction of the critical region. The x-axis
shows the nominal level of the corresponding test, while the y-axis shows
the percentage of rejection in the Monte Carlo. The dashed line corresponds
to sampling frequency of $n=1000$, the dotted line to sampling frequency
of $n=5000$ and the solid line to sampling frequency of $n=24\mbox{,}000$.}
\label{fig:test_j_m}
\end{figure}

Figure \ref{fig:hist_d} shows kernel density estimates of $U(F,k_n)_T/N_T^n$,
and Figure \ref{fig:test_d} shows the size and power of the test for disjoint
jumps. Overall the test behaves as prescribed by our asymptotic
results. Not
surprisingly, the size of the jumps have the strongest finite sample effect:
the last row of Figure \ref{fig:test_d}, corresponding to the
scenarios with
the smallest on average jumps, shows that for $n=1000$ we have slight
overrejection when the null is true (cases $c$ and $d$) and lower power when
the alternative is true (cases $j$ and $m$). The size distortion disappears
and the power converges to $1$ as we increase the sampling frequency.

Turning to the test for common jumps, Figure \ref{fig:hist_j} shows kernel
density estimates of $\log(S_n)$. The statistics are centered around
$0$ on
the samples in $\Omega_T^{(A,j)}$ (cases $j$ and $m$), as predicted from
our theoretical results. The distribution of $\log(S_n)$ on these samples
becomes more concentrated around the true value of $0$ as we increase the
frequency. On the other hand, on the samples in $\Omega_T^{(A,d)}$ (cases
$c$ and~$d$), the statistics are centered around $\log(0.5)$, and its
distribution remains nearly unchanged across the different sampling
frequencies (because for those samples $S_n$ converge to a random variable
and not a constant).

Figure \ref{fig:test_j} shows the size and power of the test for
common jumps
when we standardize $|S_n-1|$ by $V_n$. The test has overall good size with
the only exception being the cases with high intensity of arrival of small
size jumps (last row of the figure), for which even for $n=24\mbox{,}000$ we have
somewhat significant overrejection. On the other hand, from the first two
columns of Figure \ref{fig:test_j} we can see that, when using $V_n$, the
test has essentially no power against the considered alternatives. The lack
of power is explained after Theorem \ref{TTE-4}.

We next performed the test with rejection region $C'_n$ of (\ref{TE-15}),
corresponding to the truncated variance $V'_n=V_n\wedge v_n$, and we have
taken $v_n= k_n^{-0.125}\times\frac{1}{z(0.5,N_t^n)}$ where $N^n_T$ is
given by (\ref{TE-9}). The choice of $v_n$ reflects the fact that on
$\Omega_T^{(A,d)}$, $V_n$ is distributed approximately as $1/\chi^2_{N_T^n}$.
The results\vspace*{1pt} of the test with the truncated asymptotic variance are reported
on Figure \ref{fig:test_j_m}. The power against all alternatives
improves in
all cases, as seen from the first two columns of the figure. The cost
of this
is finite sample overrejection in the scenarios of frequent small
jumps, that is,
the last row on Figure \ref{fig:test_j_m}. The overrejection for cases III-j
and III-m is quite big.

Overall, we conclude that the test for disjoint jumps performs
well in finite samples and has relatively good power. The test for
common jumps should be always performed using the truncated
variance $V'_n$, and it can significantly overreject the null in
the case of jumps of small size. Finally, as confirmed by the
Monte Carlo, using coarser sampling frequencies in performing the
tests leads to larger errors in estimating the left and right
volatility. Therefore, our ability to distinguish small price and
volatility jumps worsens in such cases. As a result, on coarser
frequencies the tests will perform worse (i.e., weaker power
against alternatives and possible size distortions) when jumps are
small, for example, case III in our Monte Carlo, and there will be
little effect when jumps are bigger, for example, cases I and II
considered here.

\section{\texorpdfstring{Empirical application.}{Empirical application}}\label{sec:DATA}

Before going to the empirical application, let us mention a
crucial point. Our construction of the tests assumes that the
stochastic process is observed without error, and the Monte Carlo
in the previous section is conducted in this way. In financial
applications at very high frequencies, for example, seconds, the presence
of microstructure noise in the prices is nonnegligible. If, for
example, we have an i.i.d. noise, say with a continuous bounded
density $\phi$, then $\frac{\Delta_n}{u_n^3}\widehat{c}(k_n)_i$
converges in
probability to $\frac23 \int\phi(x)\phi(-x)\,dx$ for all $i$: so
obviously our test statistics behave in a very different way than
in our theorems for their limiting behavior in probability, not to
mention the CLTs. Intuitively, the microstructure noise will
tend to bias downwards the estimated difference between left and
right volatility, that is, a bias in favor of no common price and
volatility jumps hypothesis.

There seem to be two ways to get around the problem of
microstructure noise. One is to use a coarser frequency at which
the microstructure noise is considered as being negligible. Given
our conclusions from the Monte Carlo, this way will inevitably
sacrifice somewhat the performance of the tests when very small
jumps are involved. An alternative is to develop tests which are
robust against the noise, like using a pre-averaging preliminary
procedure for our local volatility estimators, but this will
inevitably lead to a further decrease in the rates of convergence.
Furthermore such an extension of our tests, while building on the
theoretical results here, asks for a significantly more involved
mathematical approach which goes beyond the scope of the current
paper and is thus left for future work.

In our empirical application we use one minute S\&P 500 index
futures data. The S\&P 500 index futures contract is one of the
most liquid financial instruments, and thus the microstructure
noise should be of little concern at the selected one minute
frequency. The sample period is from January 1997 till June 2007
and has 2593 trading days. We aggregate the data into business
weeks (a total of 552) and perform the tests over these periods.
Our choice for $F$ is $g(y,z)1_{\{|x|>a\}}$ with $g(y,z)$ given by
(\ref{TE-7}), and we report results for various truncation sizes
$a$. The choice of $u_n$, $k_n$ and $v_n$ is done exactly as in
the Monte Carlo study above.

%
\begin{table}
\caption{Testing for disjoint and common price and volatility
jumps for S\&P 500 index data}
\label{tb:emp_tests}
\begin{tabular*}{\textwidth}{@{\extracolsep{\fill}}lccccc@{}}
\hline
&  & \multicolumn{4}{c@{}}{\textbf{Rejection rate}} \\[-4pt]
&  & \multicolumn{4}{c@{}}{\hrulefill} \\
& \multicolumn{1}{c}{\multirow{2}{45pt}[-7pt]{\centering{\textbf{\# of weeks with
jumps}}}}
& \multicolumn{2}{c}{\textbf{Null}\mbox{${}\bolds{=}{}$}$\bolds{\Omega_T^{(A,d)}}$}
& \multicolumn{2}{c@{}}{\textbf{Null}\mbox{${}\bolds{=}{}$}$\bolds{\Omega_T^{(A,j)}}$}\\[-4pt]
& & \multicolumn{2}{c}{\hrulefill} & \multicolumn{2}{c@{}}{\hrulefill}\\
\textbf{Jump size} & &
$\bolds{\alpha=5\%}$ & $\bolds{\alpha=10\%}$ & $\bolds{\alpha=5\%}$ &
\multicolumn{1}{c@{}}{$\bolds{\alpha=10\%}$}\\
\hline
any size & 238 & $60.50\%$ & $64.71\%$ & $42.02\%$ & $51.26\%$\\
$>$0.2\% & 163 & $61.96\%$ & $65.64\%$ & $40.49\%$ & $50.31\%$\\
$>$0.3\% & \phantom{0}96 & $69.79\%$ & $70.83\%$ & $38.54\%$ & $48.96\%$\\
$>$0.4\% & \phantom{0}56 & $73.21\%$ & $73.21\%$ & $42.86\%$ & $50.00\%$\\
\hline
\end{tabular*}
\legend{Note: the test for common jumps is based on $C_n'$ in
(\ref{TE-15}).}
\end{table}

%
\begin{figure}

\includegraphics{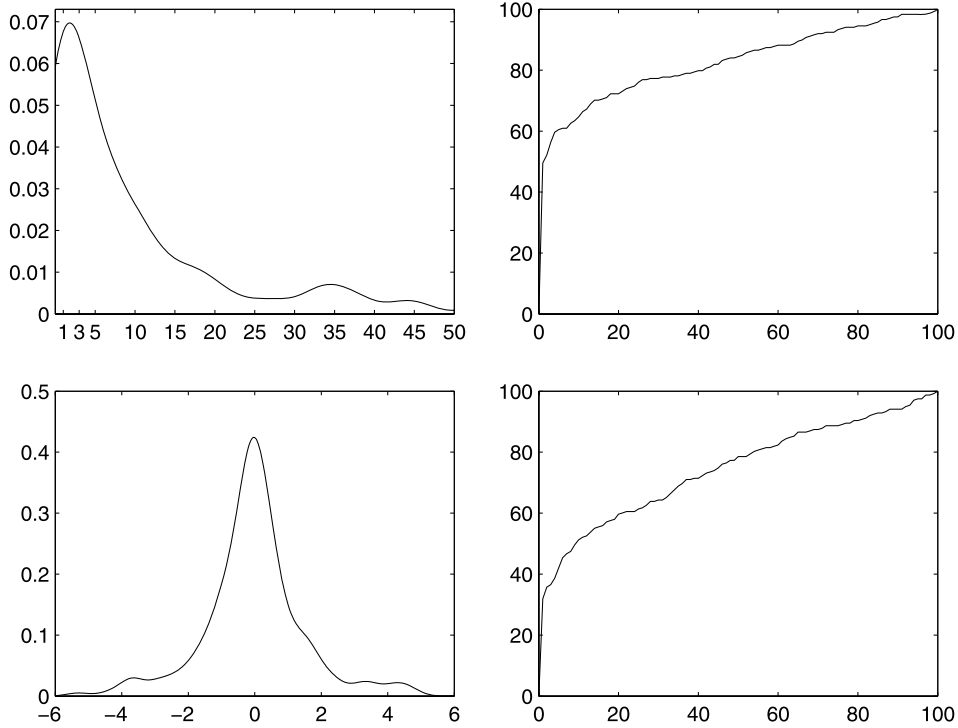}

\caption{Test results for S\&P 500 index data for truncation level
$a=0$. The
top and bottom left panels show kernel density estimates of
$U(f,g,k_n)_T/N_t^n$ and $\log(S_n)$, respectively. The top and bottom right
panels plot empirical rejection rates against nominal size of the tests for
disjoint and common jumps, respectively. The latter one is based on
$C_n'$ in
(\protect\ref{TE-15}).}
\label{fig:test_data}
\end{figure}

Table \ref{tb:emp_tests} reports the rejection rates of the two tests
(for the
conventional $5\%$ and $10\%$ significance levels) for various levels
of the
truncation size $a$, while Figure \ref{fig:test_data} plots the kernel density
estimate of the test statistics together with rejection curves of the two
tests for the case of $a=0$. The results suggest very strongly that the jumps
in the level of the S\&P 500 index are accompanied by jumps in its volatility.
This is further confirmed from Table \ref{tb:emp_decis} in which we
report the
percentage of weeks in which both tests suggest the observed path is in
$\Omega_T^{(A,j)}$, $\Omega_T^{(A,d)}$, or disagree. Based on the
results in
Table \ref{tb:emp_decis} for the weeks in which the S\&P 500 index
jumps: (1)
in approximately $40\%$ of them there is strong evidence for common
price and
volatility jumps, (2) in around $20\%$ of them there is evidence for disjoint
jumps and (3) for the rest of the weeks the tests are inconclusive.
Given our
Monte Carlo study, this last part of the sample can be explained with a lot
of small jumps for which detecting common or disjoint arrival needs even
higher frequencies.

\section{\texorpdfstring{Conclusion.}{Conclusion}}
In this paper we derive tests for deciding whether jumps in a stochastic
process are accompanied by simultaneous jumps in its volatility using only
high-frequency data of the process. Our application of the tests to S\&P
500
%
%
\begin{table}[b]
\tablewidth=270pt
\caption{Decision matrix based on the two tests for~S\&P~500~index data}
\label{tb:emp_decis}
\begin{tabular*}{\tablewidth}{@{\extracolsep{\fill}}lcc@{}}
\hline
& \textbf{Accept} $\bolds{\Omega_T^{(j)}}$ & \textbf{Reject} $\bolds{\Omega_T^{(j)}}$\\
\hline
Accept $\Omega_T^{(d)}$& $19.33\%$ & $20.17\%$\\[2pt]
Reject $\Omega_T^{(d)}$& $38.66\%$ & $21.85\%$\\
\hline
\end{tabular*}
\legend{Note: numbers based on the two tests with $5\%$ significance level and
truncation level $a=0$. The test for common jumps is based on $C_n'$ in
(\ref{TE-15}).}
\end{table}
index data indicates that most stock market jumps are associated with
volatility jumps as well.

\section{\texorpdfstring{Proofs.}{Proofs}}\label{proofs}

\subsection{\texorpdfstring{Preliminaries.}{Preliminaries}}

Under Assumptions \ref{asssumHr} and \ref{asssumKv}, both $X$ and $Z$ are It\^o
semimartingales, with (\ref{S1}) for $X$, and $Z$ has a similar
representation, in which (up to ``augmenting'' the Poisson measure
$\mu$) it is no restriction to assume that the Poisson measure is
the same. That is, we can write
%
%
\begin{eqnarray}\label{S101}
Z_{t}& =&Z_{0}+\int_{0}^{t}\widehat{b}_{s}\,ds+
\int_{0}^{t}\widehat{\sigma}_{s} \,dW_{s}+\int_{0}^{t}\widehat
{\sigma}_{s}' \,dW'_{s}\nonumber\\
&&{}
+\int_{0}^{t}\int_E\bigl(\widehat{\delta}(s,z)1_{\{|\widehat{\delta
}(t,z)|\leq1\}}\bigr)(\mu-\nu
)(ds,dz)\\
&&{} +\int_{0}^{t}\int_E\bigl(\widehat{\delta}(s,z)1_{\{|\widehat{\delta
}(t,z)|>1\}}\bigr)\mu(ds,dz),\nonumber
\end{eqnarray}
where $W'$ is another standard Brownian motion, independent of $W$.
Moreover we have $|\widehat{\delta}(\omega,t,z)|\leq\Gamma
_t(\omega)\widehat{\gamma}(z)$, where
we can always take the same process $\Gamma_t$ than in Assumption \ref{asssumHr}
for $X$, as we may do for the process $\Gamma$ showing in (\ref{S2}).
Note also that
%
%
\begin{eqnarray}\label{PR-0}
v&\leq&\frac12 \quad\Rightarrow\quad\int\bigl(\widehat{\gamma}(z)^2\wedge1\bigr)
\lambda
(dz)<\infty,\nonumber\\[-8pt]\\[-8pt]
v&>&\frac12 \quad\Rightarrow\quad\int\bigl(\widehat{\gamma}(z)^{1/v}\wedge1\bigr)
\lambda
(dz)<\infty,\qquad
\widehat{\sigma}=\widehat{\sigma}'=0.\nonumber
\end{eqnarray}

By a well-known localization procedure (see, e.g., \cite{JSEMSTAT})
it is enough to prove all theorems of Section \ref{sec:KP}, hence also of
Section \ref{sec:TS}, when in addition to the relevant Assumptions
\ref{asssumHr} and \ref{asssumKv} we have
%
%
\begin{eqnarray}\label{PR-1}
&&|b_t|+|\sigma_t|+\frac1{|\sigma_t|}+|\widehat{b}_t|+|\widehat
{\sigma}_t|+|\widehat{\sigma}
'_t|+\Gamma_t+|X_t|\nonumber\\[-8pt]\\[-8pt]
&&\qquad{}
+|Z_t|+|\overline{Z}_t|+\gamma(z)+\widehat{\gamma}(z) \leq C\nonumber
\end{eqnarray}
for some constant $C$. This additional assumption will be supposed
throughout. In the sequel, $K$ is a constant which
varies from line to line and may depend on $C$ above and also on
$r,v,\varpi$ and on the function $\gamma$ in (H-$v$), and is written
$K_q$ if
it depends on an additional parameter $q$.

Under (\ref{PR-1}), we can write $X$ as $X=X'+X''$, where
\begin{eqnarray}
X''_t &=& \cases{
\displaystyle\int_0^t\int_E\delta(s,z)(\mu-\nu)(ds,dz),&\quad if $r>1$,\vspace*{1pt}\cr
\displaystyle\int_0^t\int_E\delta(s,z)\mu(ds,dz), &\quad if $r\leq1$,}
\nonumber\\
X'_t &=& X_0+\int_0^tb'_s\,ds+\int_0^t\sigma_s \,dW_s
\nonumber\\
&& \eqntext{\mbox{where }b'_t = \cases{
\displaystyle b_t+\int_{\{|\delta(t,z)|>1\}}\delta(t,z) \lambda(dz), &\quad if
$r>1$,\vspace*{2pt}\cr
\displaystyle b_t-\int_{\{|\delta(t,z)|\leq1\}}\delta(t,z) \lambda(dz),
&\quad
if $r\leq1$.}}
\end{eqnarray}

We also need a long series of additional notation.
For each integer $m\geq1$ we denote by $(S(m,q)\dvtx q\geq1)$
the successive jump times of the counting (Poisson) process
$\mu([0,t]\times\{z\dvtx\frac1m<\gamma(z)\leq\frac1{m-1}\})$. We relabel
the two-parameter sequence $(S(m,q)\dvtx m,q\geq1)$ as a single sequence
$(T_p\dvtx p\geq1)$, which clearly exhausts the jumps of $X$.

When $m\geq1$ we denote by $\mathcal{T}_m$ the set of all $p$'s such that
$T_p=S(m',q)$ for some $q\geq1$ and $m'\in\{1,\ldots,m\}$. We set
$I(n,i)=((i-1)\Delta_n,i\Delta_n]$ and
\begin{eqnarray*}
i(n,p) &=& \mbox{the unique integer such that } T_p\in I(n,i(n,p)),\\
J(n,m) &=& \{i(n,p)\dvtx p\in\mathcal{T}_m\}, \qquad
J'(n,m) = \mathbb{N}^*\setminus J(n,m),\\
\Omega_{n,t,m} &=& \bigcap_{p\neq q, p,q\in\mathcal{T}_m}\{T_p>t\mbox{, or
$T_p>3k_n\Delta_n$ and $|T_p-T_q|>6k_n\Delta_n$}\}.
\end{eqnarray*}
We have
%
%
\begin{equation}\label{PR-22}
\lim_{n\to\infty} \mathbb{P}(\Omega_{n,t,m}) = 1.
\end{equation}

When $m\in\mathbb{N}$ we also set
%
%
\begin{eqnarray}\label{PR-23}
A_m &=& \{z\dvtx\gamma(z)\leq1/m\},\qquad \gamma_m = \int_{A_m}\gamma
(z)^r \lambda(dz),\nonumber\\
b'(m)_t &=& \cases{
\displaystyle b'_t-\int_{(A_m)^c}\delta(t,z) \lambda(dz), &\quad if $r>1$,\cr
b'_t, &\quad if $r\leq1$,}
\nonumber\\
X'(m)_t &=& X_0+\int_0^tb'(m)_s \,ds+\int_0^t\sigma_s
\,dW_s,\nonumber\\[-8pt]\\[-8pt]
Y(m)_t &=& \int_0^t\int_{(A_m)^c}\delta(s,z)\mu(ds,dz),\nonumber\\
X''(m)_t &=& \cases{
\displaystyle \int_0^t\int_{A_m}\delta(s,z)
(\mu-\nu)(ds,dz), &\quad if $r>1$,\cr
\displaystyle \int_0^t\int_{A_m}\delta(s,z)
\mu(ds,dz), &\quad if $r\leq1$,}
\nonumber\\
\overline{Y}(m) &=& X'(m)+X''(m) = X-Y(m).\nonumber
\end{eqnarray}
Note that $A_0=E$, $b'(0)=b'$, $Y(0)=0$,
$X'(0)=X'$ and $X''(0)=X''$. When $r\leq1$, we can also
define those quantities when $m=\infty$, in which case
$A_\infty=\{z\dvtx\gamma(z)=0\}$, $b'(\infty)=b'$, $Y(\infty)=X''$,
$X'(\infty)=X'$ and $X''(\infty)=0$.

Next, similar to (\ref{Z6}), we put
%
%
\begin{equation}\label{Z6-1}
\eta(k_n)_i = \frac1{k_n\Delta_n}\sum_{j=1}^{k_n}|\Delta^n_{i+j}W|^2.
\end{equation}
This notation, as well as (\ref{Z6}), is extended for convenience to the
case where \mbox{$i\leq0$}, with the convention that $\Delta^n_iY=0$ when
$i\leq0$
for any process $Y$. Finally, we set\looseness=1
\begin{eqnarray*}
\widehat{c}(k_n,p-)&=&\widehat{c}(k_n)_{i(n,p)-k_n-1},\qquad
\widehat{c}(k_n,p+)=\widehat{c}(k_n)_{i(n,p)},\\
\eta(k_n,p-) &=& \eta(k_n)_{i(n,p)-k_n-1},\qquad
\eta(k_n,p+)=\eta(k_n)_{i(n,p)},\\
\kappa(k_n,p-) &=& \sqrt{k_n} \bigl(\widehat{c}(k_n,p-)-c_{T_p-} \bigr),\qquad
\kappa(k_n,p+)=\sqrt{k_n} \bigl(\widehat{c}(k_n,p+)-c_{T_p} \bigr),\\
\kappa'(k_n,p-) &=& \sqrt{k_n} \bigl(\eta(k_n,p-)-1 \bigr),\qquad
\kappa'(k_n,p+)=\sqrt{k_n} \bigl(\eta(k_n,p+)-1 \bigr).
\end{eqnarray*}

\subsection{\texorpdfstring{Estimates.}{Estimates}}\label{ss:P0}

We proceed here by recalling or proving a number of useful estimates.
As said before, we always assume Assumptions \ref{asssumHr} and \ref{asssumKv}
and~(\ref{PR-1}).
Mostly, these estimates are conditional with respect to a possibly larger
filtration than $(\mathcal F_t)$. So we fix $m\in\mathbb{N}$, and
denote by
$\mu^{(m)}$ and $\mu'^{(m)}$ the restrictions of the measure $\mu$
to the sets $\mathbb{R}_+\times A_m$ and $\mathbb{R}_+\times
(A_m)^c$, respectively.
These are two independent Poisson measures, independent of $W$ and
$W'$ as well. We denote by $\mathcal G_m$ the $\sigma$-field generated
by the measure
$\mu'^{(m)}$, and by $(\mathcal F^{(m)}_t)$ the smallest filtration containing
$(\mathcal F_t)$ and such that $\mathcal F_0^{(m)}$ contains $\mathcal G_m$.

We set $D_m=\{(\omega,s)\dvtx\mu'^{(m)}(\omega,\{s\}\times E)=1\}$
which is also the union of the graphs of the stopping times $T_p$
for $p\in\mathcal{T}_m$. Then we define the process
\[
Z(m)_t = Z_t-\sum_{s\leq t}\Delta Z_s 1_{D_m}(s).
\]
Due to the independence of $W$, $W'$, $\mu^{(m)}$ and $\mu'^{(m)}$, the
processes $W$ and $W'$ and the measure $\mu^{(m)}$ are still Wiener processes
and a Poisson random measure, relative to the filtration $(\mathcal F^{(m)}_t)$.
Hence $X'(m)$ and $X''(m)$ are It\^o semimartingales, with the same
form as in (\ref{PR-23}) (we can replace $\mu$ and $\nu$ by $\mu^{(m)}$
and its deterministic compensator because of the presence of
$1_{A_m}$) and relative to the filtration $(\mathcal F_t^{(m)})$. In
the same way
$Z(m)$ is still of the form (\ref{S101}), driven by $W$, $W'$ and
$\mu^{(m)}$ (instead of $\mu$), relative to $(\mathcal F^{(m)}_t)$
[and up
to replacing $\widehat{b}_t$ by $\widehat{b}(m)_t=\widehat{b}_t
-\int_{(A_m)^c}\widehat{\delta}(t,z) 1_{\{|\widehat{\delta
}(t,z)|\leq1\}} \lambda(dz)$,
which is still
bounded].

\subsubsection*{\texorpdfstring{1. Estimates on $\sigma$.}{1. Estimates on $\sigma$}}
The latter property, together with (\ref{PR-1}) and classical
estimates and
the fact that $\widehat{\sigma}_t=\widehat{\sigma}'_t=0$
identically when $v>1/2$ imply that
for any $p\geq1$,
%
%
\begin{equation}\label{S200}\qquad
\mathbb{E}\Bigl({\sup_{s\leq t}}|Z(m)_{R+s}-Z(m)_R|^p\mid\mathcal
F^{(m)}_R\Bigr)\leq
\cases{
K_p t^{(p/2)\wedge1}, &\quad if $v\leq1/2$,\cr
K_p t^{(pv)\wedge1}, &\quad if $v>1/2$,}
\end{equation}
for any finite $(\mathcal F^{(m)}_t)$-stopping time $R$. Since $Z$ and
$\overline{Z}$
stay in a compact set, we have
\[
|\sigma_{t+s}-\sigma_t| \leq K(|Z_{t+s}-Z_t|+|\overline
{Z}_{t+s}-\overline{Z}_t|).
\]
Moreover, $Z_t-Z_s=Z(m)_t-Z(m)_s$ if $s<t$ and $(s,t]\cap
D_m=\varnothing$. If $R$ is a finite $(\mathcal F^{(m)}_t)$-stopping, the
set $\{(R,R+t]\cap D_m=\varnothing\}$ belongs to $\mathcal F^{(m)}_0$, so
(\ref{S2}) and (\ref{S200}) yield
%
%
\begin{equation}\label{S201}\hspace*{28pt}
\mathbb{E}\Bigl({\sup_{s\leq t}}|\sigma_{R+s}-\sigma_{R}|^p\mid\mathcal
F^{(m)}_R\Bigr)\leq
Kt^{(pv)\wedge1}\qquad
\mbox{on } \{(R,R+t]\cap D_m=\varnothing\}.
\end{equation}

\subsubsection*{\texorpdfstring{2. Estimates on $X$.}{2. Estimates on $X$}}
The following classical estimates use
(\ref{PR-1}) and $|b'(m)_t|\leq Km^{(r-1)^+}$. Below, $q>0$ and
$p\geq r$ and $i$ is an integer, possibly random but $\mathcal
F^{(m)}_0$-measurable,
and we have
%
%
\begin{eqnarray}\label{PR-4}
&&\mathbb{E}\bigl(|\Delta^n_iW|^q\mid\mathcal F^{(m)}_{(i-1)\Delta_n} \bigr)\nonumber\\
&&\qquad\leq K_q\Delta
_n^{q/2},\nonumber\\
&&\mathbb{E}\bigl(|\Delta^n_iX'(m)|^q\mid\mathcal F^{(m)}_{(i-1)\Delta_n}\bigr)\nonumber\\
&&\qquad\leq
K_q\Delta_n^{q/2}\bigl(1+\Delta_n^{q/2} m^{q(r-1)^+}\bigr),\nonumber\\
&&\mathbb{E}\bigl(|\Delta^n_iX''(m)|^p\mid\mathcal F^{(m)}_{(i-1)\Delta_n} \bigr)
\\
&&\qquad\leq
\cases{
\displaystyle\frac{K_p\Delta_n\gamma_m}{m^{p-r}} \bigl(
1+(\Delta_nm^r)^{(p-1)^+} \bigr), &\quad if $r\leq1$,\vspace*{2pt}\cr
\displaystyle\frac{K_p\Delta_n\gamma_m}{m^{p-r}} \bigl(
1+(\Delta_nm^r)^{(p-2)^+/2} \bigr), &\quad if $r>1$,}
\nonumber\\
&&\mathbb{E}\bigl( \bigl|\Delta^n_iX'(m) - \sigma_{(i-1)\Delta_n}\Delta^n_iW \bigr|^q
\mid\mathcal F^{(m)}_{(i-1)\Delta_n} \bigr)\nonumber\\
&&\qquad\leq
K \bigl(\Delta_n^{q/2+(qv)\wedge 1}
+ \Delta_n^q m^{q(r-1)^+} \bigr)\nonumber\\
\eqntext{\mbox{on the set } \{I(n,i)\cap D_m=\varnothing\}.}
\end{eqnarray}

Next, we also have for $p\geq r$,
%
%
\begin{equation}\label{PR-5}
\mathbb{E}\bigl(|\Delta^n_iX''(m)|^p\wedge
u_n^p\mid\mathcal F^{(m)}_{(i-1)\Delta_n} \bigr) \leq Ku_n^{p-r}\Delta
_n\gamma_m.
\end{equation}
These estimates hold when $m=0$ as well [in which case
$\mathcal F^{(0)}_t=\mathcal F_t$ and $i$ is not random, and $Y(0)=0$]. In
particular, in
this case we deduce
%
%
\begin{equation}\label{PR-401}
\mathbb{E}\bigl(|\Delta^n_iX|^2\mid\mathcal F_{(i-1)\Delta_n} \bigr) \leq
K\Delta_n.
\end{equation}

Next, with any measurable subset $A$ of $E$ we consider the increasing
process $G(A)_t=\int_0^t\int_A\gamma(z)\mu(ds,dz)$. This process is
infinite for all $t>0$ if $\int_A\gamma(z)\times\lambda(dz)=\infty$, and otherwise
is a L\'evy process, and known estimates on L\'evy processes yield
for all $q>0$,
%
%
\begin{equation}\label{PR-2031}
\mathbb{E}((G(A)_t)^q) \leq K_q \biggl(t\int_A\gamma(z)^q\lambda(dz)+
\biggl(t\int_A\gamma(z)\lambda(dz) \biggr)^{q\vee1} \biggr).
\end{equation}
[Since $\gamma$ is bounded, when $q\leq1$ the right-hand side above
is smaller
than $K_qt\int_A\gamma(z)^q\lambda(dz)$.] Since $|\Delta
^n_iY(m)|\leq\Delta^n_iG(A_m^c)$,
we deduce (for $i\geq1$ not random)
%
%
\begin{equation}\label{PR-2030}
q\geq r \quad\Rightarrow\quad
\mathbb{E}\bigl(|\Delta^n_iY(m)^n_i|^q\mid\mathcal F_{(i-1)\Delta_n} \bigr)
\leq
K_q \bigl(\Delta_n+\bigl(\Delta_nm^{(r-1)^+}\bigr)^{q\vee1}
\bigr).\hspace*{-35pt}
\end{equation}

\subsubsection*{\texorpdfstring{3. Estimates on $\widehat{c}(k_n)_i$.}{3. Estimates
on $\widehat{c}(k_n)_i$}} Below, $i\geq1$ is a
nonrandom integer. First (\ref{PR-401}) yields
%
%
\begin{equation}\label{PR-402}
\mathbb{E}(\widehat{c}(k_n)_i\mid\mathcal F_{i\Delta_n} ) \leq K.
\end{equation}
We need also estimates on the difference $\widehat{c}(k_n)_i-c_t$ for
suitable times $t$. If $S$ is a $\mathcal F^{(m)}_0$-measurable
positive finite time and $i\geq1$ an $\mathcal F^{(m)}_0$-measurable random
integer, the sets
\begin{eqnarray*}
\Omega(m,n,S,i)_+ &=& \bigl\{(i-1)\Delta_n<S\leq i\Delta_n,
\bigl(S,S+(k_n+1)\Delta_n\bigr]\cap D_m=\varnothing\bigr\},\\
\Omega(m,n,S,i)_- &=& \bigl\{(i-1)\Delta_n<S\leq i\Delta_n,
\bigl(S-(k_n+2)\Delta_n,S\bigr)\cap D_m=\varnothing\bigr\}
\end{eqnarray*}
are $\mathcal F^{(m)}_0$-measurable, and we have
\begin{lem}\label{L01} Assume Assumptions \ref{asssumHr} and \ref{asssumKv} and (\ref{PR-1}).
Let $q=1$ or $q=2$, and assume (\ref{Z5}) with also
%
%
\begin{eqnarray}\label{PR-4031}
q&=&1\quad
\Rightarrow\quad\rho<\frac{2v}{1+2v}\wedge\bigl(2\varpi(2-r)\bigr),\nonumber\\[-8pt]\\[-8pt]
q&=&2 \quad\Rightarrow\quad\rho<\frac{(2v)\wedge1}{1+(2v)\wedge
1}\wedge
\bigl(\varpi(4-r)-1 \bigr).\nonumber
\end{eqnarray}
Then there is a sequence $\alpha_n(q)\to0$ such that,
for $m\geq0$ and any $\mathcal F^{(m)}_0$-measurable variables $S$ and $i$
as above, we have
%
%
\begin{eqnarray}\label{PR-403}
&&\mathbb{E}\bigl(|\widehat{c}(k_n)_i-c_S \eta(k_n)_i|^q\mid\mathcal
F^{(m)}_S \bigr) \nonumber\\
&&\qquad\leq
\frac{K_m \alpha_n(q)}{k_n^{q/2}}
\qquad\mbox{on } \Omega(m,n,S,i)_+,\nonumber\\[-8pt]\\[-8pt]
&&\mathbb{E}\bigl(|\widehat{c}(k_n)_{i-k_n-1}-c_{S-} \eta
(k_n)_{i-k_n-1}|^q\mid
\mathcal F^{(m)}_{(i-k_n-1)\Delta_n} \bigr)\nonumber\\
&&\qquad\leq\frac{K_m \alpha_n(q)}{k_n^{q/2}}\qquad
\mbox{on } \Omega(m,n,S,i)_-\nonumber
\end{eqnarray}
and also
%
%
\begin{eqnarray}\label{PR-4030}
\mathbb{E}\bigl(|\widehat{c}(k_n)_i-c_S|^q\mid\mathcal F^{(m)}_S \bigr) &\leq&
\frac{K_m}{k_n^{q/2}}\qquad
\mbox{on } \Omega(m,n,S,i)_+,\hspace*{-28pt}\nonumber\\[-8pt]\\[-8pt]
\mathbb{E}\bigl(|\widehat{c}(k_n)_{i-k_n-1}-c_{S-}|^q\mid
\mathcal F^{(m)}_{(i-k_n-1)\Delta_n} \bigr)&\leq&\frac{K_m}{k_n^{q/2}}\qquad
\mbox{on } \Omega(m,n,S,i)_-.\hspace*{-28pt}\nonumber
\end{eqnarray}

Moreover, as soon as $r<2$, and under (\ref{Z5}) only, we have
%
%
\begin{eqnarray}\label{PR-4033}
\widehat{c}(k_n)_i&\stackrel{\mathbb{P}}{\longrightarrow}&c_S \qquad\mbox
{on } \Omega(m,n,S,i)_+,\nonumber\\[-8pt]\\[-8pt]
\widehat{c}(k_n)_{i-k_n-1}&\stackrel{\mathbb{P}}{\longrightarrow}& c_{S-}
\qquad\mbox{on } \Omega(m,n,S,i)_-.\nonumber
\end{eqnarray}
\end{lem}
\begin{pf}
We will prove, for example, the
second claims of (\ref{PR-403}), (\ref{PR-2030}) and (\ref{PR-4033}) (the
first ones are slightly easier). On the set $\Omega(m,n,S,i)_-$ the variable
$\widehat{c}(k_n)_{i-k_n-1}$ is equal to the variable $\widehat
{c}'(k_n)_{i-k_n-1}$
associated in the same way with the process $\overline{Y}(m)$.

The following estimate, for all
$x,y,z\in\mathbb{R}$, $u>0$, $w>0$, is straightforward:
\begin{eqnarray*}
&&\bigl||x+y+z|^21_{\{|x+y+z|\leq u\}}-x^2 \bigr|^q\\
&&\qquad\leq
K_q \biggl((y\wedge u)^{2q}+z^{2q}+|x|^q(|y|\wedge u)^q
+|x|^q|z|^q+\frac{|x|^{(2+w)q}}{u^{wq}}\biggr).
\end{eqnarray*}
This will be applied with $x=\sigma_{(j-1)\Delta_n}\Delta^n_jW$ and
$y=\Delta
^n_jX''(m)$
and $z=\Delta^n_jX'(m)-\sigma_{(j-1)\Delta_n}\Delta^n_jW$ [so
$\Delta^n_j\overline{Y}
(m)=x+y+z$],
and $u=u_n$ and $w$ such that $w(1-2\varpi)\geq2$, and when $j=i-k_n-1,i-k_n,
\ldots,i-1$: using H\"older's inequality, we deduce
from (\ref{PR-4}) and (\ref{PR-5}) and the boundedness of
$\sigma_t$, and after some calculation, that in this case
\begin{eqnarray*}
&&\mathbb{E}\bigl( \bigl|(\Delta^n_j\overline{Y}(m))^2 1_{\{|\Delta
^n_j\overline{Y}(m)|\leq u_n\}}
-c_{(j-1)\Delta_n}(\Delta^n_jW)^2 \bigr|^q\mid\mathcal
F^{(m)}_{(i-k_n-1)\Delta
_n} \bigr)\\
&&\qquad\leq
K_{m,\theta} \bigl(\Delta_n^{1+(2q-r)\varpi}+\Delta_n^{q+(qv)\wedge
\theta} \bigr)
\end{eqnarray*}
for any $\theta\in(0,1)$, on the set $\Omega(m,n,S,i)_-$, because
$I(n,j)\cap D_m=\varnothing$.

Next, we write $|c_{(j-1)\Delta_n}-c_{S-}|\leq|c_{(j-1)\Delta
_n}-c_{j\Delta_n}|
+|c_{j\Delta_n}-c_{S-}|$, and we apply (\ref{S201}) and (\ref{PR-4}) and
either H\"older's inequality plus the boundedness of $\sigma_t$,
or successive conditioning, to get, for $j$ and $\theta$ as above,
\begin{eqnarray*}
&&\mathbb{E}\bigl(\bigl|c_{(j-1)\Delta_n}-c_{S-}\bigr|^q (\Delta^n_jW)^{2q}
\mid\mathcal F^{(m)}_{(i-k_n-1)\Delta_n} \bigr)\\
&&\qquad\leq
K_\theta\bigl(\Delta_n^q(k_n\Delta_n)^{(qv)\wedge1}+\Delta
_n^{q+(qv)\wedge\theta
} \bigr).
\end{eqnarray*}
These estimates, together with the definition of
$\widehat{c}'(k_n)_{i-k_n-1}$ and $\eta(k_n)_{i-k_n-1}$, yield
\[
\mathbb{E}\bigl(|\widehat{c}'(k_n)_{i-k_n-1}-c_{S-} \eta
(k_n)_{i-k_n-1}|^q\mid
\mathcal F^{(m)}_{(i-k_n-1)\Delta_n} \bigr) \leq K_{m,\theta} a(q)_n
\]
on the set $\Omega(m,n,S,i)_-$, where $a(q)_n=
\Delta_n^{1+(2q-r)\varpi-q}+\Delta_n^{((qv)\wedge1)(1-\rho)}
+\Delta_n^{(qv)\wedge\theta}$. Then (\ref{Z5}) and a proper choice of
$\theta$ show that $a(q)_nk_n^{q/2}\to0$ for $q=1,2$, under
(\ref{PR-2031}), and $a(1)_n\to0$ as soon as $r<2$. This in particular
gives the second part of (\ref{PR-403}).

Finally (\ref{PR-4030}) and (\ref{PR-4033}) follow from the above,
from the
boundedness of the process $c_t$, and from the following property:
if $R$ is any $(\mathcal F^{(m)}_t)$-stopping time and $i\Delta_n\geq
R$, then
\[
\mathbb{E}\bigl(\bigl(\eta(k_n)_i-1\bigr)^2\mid\mathcal F^{(m)}_R \bigr) = 2/k_n.
\]
This readily follows from the fact that $\eta(k_n)_i$ is
independent of $\mathcal F^{(m)}_R$ and given by (\ref{Z6-1}).
\end{pf}

\subsection{\texorpdfstring{The stable convergence of
$\widehat{c}(k_n)_i$.}{The stable convergence of $\widehat{c}(k_n)_i$}}

From now on, the integer $w\geq2$ is fixed. The aim of this subsection
is to
prove the following stable convergence:
\begin{prop}\label{PPR-1} As soon as Assumptions \ref{asssumHr}, \ref{asssumKv}, (\ref{PR-1})
and (\ref{PR-4031}) for $q=1$ hold, the sequence of variables
%
%
\begin{equation}\label{PR-9}
(\kappa(k_n,p-),\kappa(k_n,p+),\kappa(wk_n,p-),\kappa(wk_n,p+)
)_{p\geq1}
\end{equation}
converges stably in law as $n\to\infty$ (for the product topology on
$\mathbb{R}^{\mathbb{N}^*}$) to
%
%
\begin{eqnarray}\label{PR-10}
&&\Biggl(c_{T_p-}\sqrt{2} V_p^-,c_{T_p}\sqrt{2} V_p^+,
c_{T_p-}\sqrt{\frac2w}\bigl(V_p^-+\sqrt{w-1}
V_p'^-\bigr),\nonumber\\[-8pt]\\[-8pt]
&&\hspace*{124.26pt}c_{T_p}\sqrt{\frac2w}\bigl(V_p^++\sqrt{w-1} V_p'^+\bigr)
\Biggr)_{p\geq1},\nonumber
\end{eqnarray}
where the variables $V_p^-,V_p^+,V'^-_p,V'^+_p$ are defined on an
extension of the original space $(\Omega,\mathcal F,\mathbb{P})$ and
are all independent
and $\mathcal N(0,1)$-distributed, and independent of $\mathcal F$.
\end{prop}
\begin{pf}
\textit{Step} 1. It is enough to prove the
convergence of any finite sub-family of indices $p$. In other words,
instead of considering the infinite sequence indexed by $p\geq1$
in (\ref{PR-9}) and (\ref{PR-10}), we can fix an arbitrarily large
integer $P$ and consider the families indexed by $p\in\{1,\ldots,P\}$.
All $p$ smaller than $P$ are in some $\mathcal{T}_m$, and we consider
the set
$\Omega_n$ on which for any $p\leq P$ and any $q\in\mathcal{T}$ we have
$T_p>3k_n\Delta_n$ and $|T_p-T_q|>6k_n$. Obviously,
$\mathbb{P}(\Omega_n)\to1$ as $n\to\infty$.

Now we will apply Lemma \ref{L01} with $S=T_p$ for $p\leq P$, and
$i=i(n,p)$: then $S$ and $i$ are $\mathcal F^{(m)}_0$-measurable, and
the set
$\Omega_n$ is included into both $\Omega(m,n,T_p,i(n,p))_+$ and
$\Omega(m,n,T_p,i(n,p))_-$. Since $\mathbb{P}(\Omega_n)\to1$, we
deduce from this
lemma that
%
%
\begin{eqnarray}\label{PR-3}
\sqrt{k_n} \bigl(\widehat{c}(k_n,p-)-c_{T_p-}\eta(k_n,p-) \bigr) &\stackrel
{\mathbb{P}}{\longrightarrow}&0,\nonumber\\[-8pt]\\[-8pt]
\sqrt{k_n} \bigl(\widehat{c}(k_n,p+)-c_{T_p}\eta(k_n,p+) \bigr) &\stackrel
{\mathbb{P}}{\longrightarrow}&0.\nonumber
\end{eqnarray}

\textit{Step} 2. Now we set
%
%
\begin{eqnarray}\label{PR-17}\qquad
\chi^n &=& (\kappa'(k_n,p-),\kappa'(k_n,p+),\kappa'(wk_n,p-),\kappa'(wk_n,p+)
)_{1\leq p\leq P},\nonumber\\
\chi &=& \Biggl(\sqrt{2} V_p^-,\sqrt{2} V_p^+,
\sqrt{\frac2w}\bigl(V_p^-+\sqrt{w-1} V_p'^-\bigr),\\
&&\hspace*{81.43pt}\sqrt{\frac2w}\bigl(V_p^++\sqrt{w-1} V_p'^+\bigr) \Biggr)_{1\leq p\leq P}.\nonumber
\end{eqnarray}
By (\ref{PR-3}), we are left to prove that the variables
$\chi^n$ stably converge in law to $\chi$. Taking into account that
$\chi$ is independent of $\mathcal F$, this amounts to proving
%
%
\begin{equation}\label{PR-12}
\mathbb{E}(U f(\chi^n)) \to\mathbb{E}(U) \mathbb{E}(f(\chi)),
\end{equation}
where $U$ is any bounded $\mathcal F$-measurable variable,
and $f$ is continuous bounded.

In fact, if $(\mathcal G^{(m)}_t)$ denotes the smallest filtration to
which $W$
is adapted and such that $\mathcal G_m\subset\mathcal G^{(m)}_0$, each
$\chi^n$ is
$\mathcal G^{(m)}_\infty$-measurable. So, up to substituting $U$ with
$E(U\mid\mathcal G_\infty)$ above, it is clearly enough to prove
(\ref{PR-12})
when $U$ is $\mathcal G^{(m)}_\infty$-measurable.

\textit{Step} 3. We introduce some further notation: first the
set $F_n=\bigcup_{1\leq p\leq P}((T_p-(wk_n+1)\Delta
_n)^+,T_p+(wk_n+1)\Delta_n]$,
which is a random $\mathcal G^{(m)}_0$-measurable set, and second the processes
\[
\overline{W}{}^n_t = \int_0^t1_{F_n}(s) \,dW_s,\qquad
\overline{W}{}'^n = W-\overline{W}{}^n
\]
[those are well defined because $W$ is a $(\mathcal
G^{(m)}_t)$-Brownian motion].
The $\sigma$-fields $\mathcal H_n$ generated by $\mathcal G^{(m)}_0$
and all variables
$\overline{W}{}'^n_t$ increase with $n$, and $\bigvee_n\mathcal
H_n=\mathcal G^{(m)}_\infty$.
Therefore it is enough to prove (\ref{PR-12}) when $U$ is $\mathcal
H_q$-measurable
for some $q$: to see this, let $U$ be $\mathcal G^{(m)}_\infty
$-measurable; set
$U_q=\mathbb{E}(U\mid\mathcal H_q)$; if (\ref{PR-12}) holds for
each~$U_q$, it also
holds for $U$ because $U_q\to U$ in $\mathbb{L}^1(\mathbb{P})$.

The set $\Omega_n$ of Step 1
is $\mathcal G_m$-measurable, hence $\mathcal H_q$-measurable for all $q$.
Since $\mathbb{P}(\Omega_{n})\to1$ it is enough to prove that for
any bounded
$\mathcal H_q$-measurable variable $U$,
%
%
\begin{equation}\label{PR-16}
\mathbb{E}(U 1_{\Omega_n} f(\chi^n) ) \to\mathbb{E}(U) \mathbb
{E}(f(\chi)).
\end{equation}

\textit{Step} 4. We introduce a double sequence $(N(p,i)\dvtx p,i\geq1)$
of i.i.d. $\mathcal N(0,1)$ variables on some auxiliary probability
space. Then,
define the variables
\begin{eqnarray*}
\zeta(k_n,p-) &=& \frac1{\sqrt{k_n}}\sum_{i=1}^{k_n}\bigl(N(p,i)^2-1\bigr),\\
\zeta(k_n,p+) &=& \frac1{\sqrt{k_n}}\sum
_{i=wk_n+1}^{(w+1)k_n}\bigl(N(p,i)^2-1\bigr),\\
\zeta(wk_n,p-) &=& \frac1{\sqrt{wk_n}}\sum_{i=1}^{wk_n}\bigl(N(p,i)^2-1\bigr),\\
\zeta(wk_n,p+) &=& \frac1{\sqrt{wk_n}}\sum_{i=wk_n+1}^{2wk_n}\bigl(N(p,i)^2-1\bigr).
\end{eqnarray*}
Observe that in restriction to the set $\Omega_n$ the variable $\chi^n$
involves increments of $W$ which are different for different values of $p$,
and are increments of the process $\overline{W}{}^n$ above, which is
independent of
$\overline{W}{}'^n$. Therefore if $q$ is fixed, for any $n\geq q$ and
in restriction to the $\mathcal H_q$-measurable set $\Omega_n$, the
$\mathcal H_q$-conditional
distribution of the variable $\chi^n$ of (\ref{PR-17}) is exactly the
law of
\[
\zeta_n = (\zeta(k_n,p-),\zeta(k_n,p+),\zeta(wk_n,p-),\zeta
(wk_n,p+) )
_{1\leq p\leq P}.
\]
This means that the left-hand side of (\ref{PR-16}) for $n\geq q$ is equal
to $\mathbb{E}(U1_{\Omega_n})\times\break \mathbb{E}(f(\zeta_n))$.

At this stage, we see that (\ref{PR-16}) amounts to proving that
$\zeta_n$ converges in law to the variable $\chi$ given in the second
half of (\ref{PR-17}). This is an obvious consequence of the $4P$-dimensional
ordinary central limit theorem.
\end{pf}

\subsection{\texorpdfstring{Proof of Theorem \protect\ref{TKP-1}.}{Proof of Theorem 3.1}}

1. As stated before, we assume Assumptions \ref{asssumHr} and \ref{asssumKv} and (\ref{PR-1}).
If $m\geq1$ and $J(n,m,t) = J'(n,m)\cap\{k_n+1,k_n+2,\ldots,
[t/\Delta_n]-k_n\}$ and $\mathcal{T}_m(n,t)=\{p\in\mathcal
{T}_m\dvtx T_p\leq\Delta_n[t/\Delta
_n]\}$,
we have
%
%
\begin{eqnarray}\label{PR-25}
t\leq T \quad\Rightarrow\quad U(F,k_n)_t =
\widetilde{U}^n(m)_t+\overline{U}{}^n(m)_t\nonumber\\
\eqntext{\mbox{on the set $\Omega_{n,T,m}$, where}}\\
\\[-18pt]
\eqntext{\displaystyle\widetilde{U}^n(m)_t=\sum_{p\in\mathcal{T}_m(n,t)}F\bigl(\Delta^n_{i(n,p)}X,
\widehat{c}(k_n,p-),\widehat{c}(k_n,p+)\bigr)1_{\{\Delta
^n_{i(n,p)}X|>u_n\}}}
\\
\eqntext{\displaystyle\overline{U}{}^n(m)_t=\sum_{i\in J(n,m,t)} F(\Delta^n_i\overline
{Y}(m),\widehat{c}(k_n)_{i-k_n-1},\widehat{c}(k_n)_i)
1_{\{|\Delta^n_i\overline{Y}(m)|>u_n\}}.}
\end{eqnarray}
The sum defining $\widetilde{U}^n(m)_t$ has a bounded number of
summands, as $n$
varies. We also have for $p\in\mathcal{T}_m$,
%
%
\begin{eqnarray}\label{PR-2502}
\Delta^n_{i(n,p)}X&\to&\Delta X_{T_p},\qquad \mathbb{P}\bigl(\bigl|\Delta
^n_{i(n,p)}\overline{Y}
(m)\bigr|>u_n/2\bigr)\to0,\nonumber\\[-8pt]\\[-8pt]
\Delta^n_{i(n,p)}X&=&\Delta X_{T_p}+\Delta^n_{i(n,p)}\overline{Y}(m)
\qquad\mbox{on }
\Omega_{n,t,m}\nonumber
\end{eqnarray}
[use (\ref{PR-4}) and $\sqrt{\Delta_n}/u_n\to0$ for the second
property]. We have
\mbox{$\mathbb{P}(\Delta X_{T_p}\in R)=1$} by (\ref{Z701})\vspace*{-1pt} and $F$ is
continuous on
$R\times\mathbb{R}^{*2}_+$. Since
$\widehat{c}(k_n,p-)\stackrel{\mathbb{P}}{\longrightarrow}c_{T_p-}$
and $\widehat{c}(k_n,p+)\stackrel{\mathbb{P}}{\longrightarrow
}c_{T_p}$ on
$\Omega_{n,t,m}\cap\{T_p\leq t\}$ [use (\ref{PR-4033}) with
$S=T_p$], the
$p$th summand in $\widetilde{U}^n(m)_t$ converges to $F(\Delta X_{T_p},
c_{T_p-},c_{T_p})1_{\{\Delta X_{T_p}\neq0\}} 1_{\{T_p\leq t\}}$
in probability. Therefore we have the following convergence in
probability for the Skorokhod topology:
%
%
\begin{equation}\label{PR-2501}
\widetilde{U}^n(m)_t \stackrel{\mathbb{P}}{\longrightarrow
}\widetilde{U}(m)_t = \sum_{p\in\mathcal{T}_m}F(\Delta X_{T_p},
c_{T_p-},c_{T_p}) 1_{\{T_p\leq t\}}.
\end{equation}

2. Next, we show the result in case (a). Pick $m>2/\varepsilon$. Since
$|\Delta\overline{Y}(m)_s|\leq1/m$, for any $t>0$ we have $|\Delta
^n_i\overline{Y}(m)|\leq
2/m$ for
all $i\leq[t/\Delta_n]$, on a set $\Omega^n_t$ whose probability
goes to $1$.
On $\Omega^n_t$ we have $\overline{U}{}^n(m)_s=0$ for all $s\leq t$,
because of the
property of $F$, which also implies $\widetilde{U}(m)=U(F)$
identically. Then
the result readily follows from (\ref{PR-2501}).

3. Next, we show the result in case (b). The notation (\ref{PR-23})
is also valid for $m=\infty$, and (\ref{PR-2501}) holds for $m=\infty$
(the right-hand side is a finite sum) and $\widetilde{U}(\infty
)=U(F)$. Since
$\overline{Y}(\infty)=X'(\infty)$, it follows from the second part
of (\ref{PR-4})
(which also holds with $m=\infty$ when $r=0$) that
$\mathbb{P}(\Delta^n_i\overline{Y}(\infty)|>u_n)\leq K_q\Delta_n^{q/2}
u_n^{-q}$, which is smaller than $K\Delta_n^2$ if $q=\frac4
{1-2\varpi}$. So Borel--Cantelli lemma yields that, for each $t$, we have
$|\Delta^n_i\overline{Y}(\infty)|\leq u_n$ for all $i\leq[t/\Delta
_n]$, hence $\overline{U}{}
^n(\infty)_s=0$
for $s\leq t$, when $n$ is large enough. We then conclude as above.

4. It remains to consider the case (c). First, $|F(\Delta X_s,c_{s-},c_s)|
\leq K|\Delta X_s|^r$ as soon as $|\Delta X_s|\leq\varepsilon$
(recall that $c_s$ is
bounded). Since $\sum_{s\leq t}|\Delta X_s|^r<\infty$ a.s.
for all~$t$, whereas $|\Delta X_s|\leq1/m$ when $s$ differs from all $T_p$ for
$p\in\mathcal{T}_m$, we deduce from the dominated convergence theorem that
$\widetilde{U}(m)\to U(F)$ a.s., locally uniformly in time as $m\to
\infty$.
Therefore by (\ref{PR-2501}) it remains to prove that for all $t>0$,
%
%
\begin{equation}\label{PR-26}
\eta>0 \quad\Rightarrow\quad
\lim_{m\to\infty} \limsup_{n\to\infty}
\mathbb{P}\Bigl({\sup_{s\leq t}}|\overline{U}{}^n(m)_s|>\eta\Bigr) = 0.
\end{equation}

On the one hand, as in the previous step we deduce from (\ref{PR-4}) and
from $|\Delta X''(m)_s|\leq1/m$ that, if $m>4/\varepsilon$, we have
$|\Delta^n_iX'(m)|\leq u_n/2$ and $|\Delta^n_iX''(m)|\leq\varepsilon
/2$ for all $i\leq
[t/\Delta_n]$,
when $n$ is large enough. On the other hand, our assumption on $F$
yields that
if $|x|\leq u_n/2$ and $|x'|\leq\varepsilon/2$, then
$|F(x+x',y,z)|1_{\{
|x+x'|>u_n\}}
\leq K|x'|^r(1+y+z)$ as soon as $u_n\leq\varepsilon$. Hence for any
given $t$,
and outside a set $\Omega'_{n,t,m}$ satisfying $\mathbb{P}(\Omega
'_{n,t,m})\to1$ as
$n\to\infty$, we have $|\overline{U}{}^n(m)_s|\leq H^n(m)_t$ for all
$s\leq t$, where
\[
H^n(m)_t =
K\sum_{i=k_n+1}^{[t/\Delta_n]-k_n}|\Delta^n_iX''(m)|^r \bigl(1+\widehat
{c}(k_n)_{i-k_n-1}+\widehat{c}(k_n)_i\bigr).
\]
Therefore we are left to show that for all $t$,
%
%
\begin{equation}\label{PR-27}
\lim_{m\to\infty} \sup_n \mathbb{E}(H^n(m)_t ) = 0.
\end{equation}

The estimates (\ref{PR-2030}) and (\ref{PR-402}) and successive conditioning
yield that
\[
\mathbb{E}\bigl(|\Delta^n_iX''(m)|^r \bigl(1+\widehat
{c}(k_n)_{i-k_n-1}+\widehat{c}(k_n)_i\bigr)
\bigr) \leq K \Delta_n \gamma_m.
\]
Since $\gamma_m\to0$ as $m\to\infty$, we deduce (\ref{PR-27}) and
Theorem \ref{TKP-1} is proved.

\subsection{\texorpdfstring{Proof of Theorem \protect\ref{TKP-2}.}{Proof of Theorem 3.2}}
We need many steps, and as before we
assume Assumptions \ref{asssumHr} and \ref{asssumKv}, and also (\ref{PR-1}).

\textit{Step} 1. We use the notation (\ref{PR-25}) of the previous
proof when we deal with $k_n$ and write instead $\widetilde{U}'^n(m)$ and
$\overline{U}{}'^n(m)$ when we deal with $wk_n$. We also use $\widetilde
{U}(m)_t$, as defined
in (\ref{PR-2501}), and
\[
\widehat{U}^n(m)_t = \overline{U}{}^n(m)_t-\sum_{p\geq1, p\notin
\mathcal{T}_m}
F(\Delta X_{T_p},c_{T_p-},c_{T_p})1_{\{T_p\leq t\}}
\]
and $\widehat{U}'(m)$ is the same with $\overline{U}{}'^n(m)$ instead
of $\overline{U}
{}^n(m)$. We have
\[
\widetilde{U}^n(m)_t-\widetilde{U}(m)_t = \sum_{p\in\mathcal
{T}_m}\zeta^n_p,\qquad
\widetilde{U}'^n(m)_t-\widetilde{U}(m)_t = \sum_{p\in\mathcal
{T}_m}\zeta'^n_p,
\]
where
\begin{eqnarray*}
\zeta^n_p &=& F\bigl(\Delta^n_{i(n,p)}X,\widehat{c}(k_n,p-),\widehat{c}(k_n,p+)\bigr)
1_{\{\Delta^n_{i(n,p)}X|>u_n\}} 1_{\{T_p\leq\Delta_n[t/\Delta_n]\}
}\\
&&{} -F(\Delta X_{T_p},c_{T_p-},c_{T_p})
1_{\{\Delta X_{T_p}\neq0\}} 1_{\{T_p\leq t\}},\\
\zeta'^n_p &=& F\bigl(\Delta^n_{i(n,p)}X,\widehat{c}(wk_n,p-),\widehat{c}(wk_n,p+)\bigr)
1_{\{\Delta^n_{i(n,p)}X|>u_n\}} 1_{\{T_p\leq\Delta_n[t/\Delta_n]\}
}\\
&&{} - F(\Delta X_{T_p},c_{T_p-},c_{T_p})
1_{\{\Delta X_{T_p}\neq0\}} 1_{\{T_p\leq t\}}.
\end{eqnarray*}

We also set
%
%
\begin{eqnarray}\label{PR-321}
\overline{\zeta}{}^n_p&=& \bigl(F'_2(\Delta X_{T_p},c_{T_p-},c_{T_p})
\kappa(k_n,p-)\nonumber\\
&&\hspace*{1.4pt}{} +F'_3(\Delta X_{T_p},c_{T_p-},c_{T_p})\kappa
(k_n,p+) \bigr)
1_{\{\Delta X_{T_p}\neq0\}},\nonumber\\[-8pt]\\[-8pt]
\overline{\zeta}{}'^n_p&=& \bigl(F'_2(\Delta X_{T_p},c_{T_p-},c_{T_p})
\kappa(wk_n,p-)\nonumber\\
&&\hspace*{1.4pt}{} +F'_3(\Delta X_{T_p},c_{T_p-},c_{T_p})\kappa
(wk_n,p+) \bigr)
1_{\{\Delta X_{T_p}\neq0\}}.\nonumber
\end{eqnarray}

\textit{Step} 2. In this step we prove that
%
%
\begin{eqnarray}\label{PR-322}
&&\bigl(\sqrt{k_n} \bigl(\widetilde{U}^n(m)-\widetilde{U}(m) \bigr),
\sqrt{k_n} \bigl(\widetilde{U}'^n(m)-\widetilde{U}(m) \bigr)
\bigr)\nonumber\\[-8pt]\\[-8pt]
&&\quad\mbox{$\stackrel{\mathcal L-\mathrm{s}}{\Longrightarrow}\quad$}
\biggl(\mathcal U(m),\frac1w \bigl(\mathcal U(m)+\sqrt{w-1} \mathcal U'(m) \bigr)
\biggr)\nonumber
\end{eqnarray}
(stable functional convergence in law) where $\mathcal U(m)$ and
$\mathcal U'(m)$ are
as described in (\ref{Z-12}), except that the sum is taken over the
$p\in\mathcal{T}_m$ only. By Proposition \ref{PPR-1}, we have
\[
\sum_{p\in\mathcal{T}_m} (\overline{\zeta}{}^n_p,\overline{\zeta
}{}'^n_p ) 1_{\{T_p\leq t\}}
\quad\stackrel{\mathcal L-\mathrm{s}}{\Longrightarrow}\quad \biggl(\mathcal
U(m)_t,\frac1{\sqrt{w}}
\bigl(\mathcal U(m)_t+\sqrt{w-1} \mathcal U'(m)_t \bigr) \biggr);
\]
note the normalization in $\kappa(wk_n,p\pm)$ is by $\sqrt{wk_n}$. Hence
proving (\ref{PR-322}) shows that for each $p\in\mathcal{T}_m$ we have
%
%
\begin{equation}\label{PR-320}
\sqrt{k_n} \zeta^n_p-\overline{\zeta}{}^n_p \stackrel{\mathbb
{P}}{\longrightarrow}0,\qquad
\sqrt{wk_n} \zeta'^n_p-\overline{\zeta}{}'^n_p \stackrel{\mathbb
{P}}{\longrightarrow}0,
\end{equation}
in restriction to each set $\{T_p\leq t\}$. We will prove, for example,
the first property. We have $\mathbb{P}(\Delta_n[t/\Delta_n]<T_p\leq
t)\to0$ and
(\ref{PR-22}) and (\ref{PR-2502}), implying that the set
$\{|\Delta^n_{i(n,p)}X|>u_n\}$ converges in probability to the set
$\{\Delta X_{T_p}\neq0\}$. Therefore it is enough to show that
\begin{eqnarray*}
&&\sqrt{k_n} \bigl(F\bigl(\Delta^n_{i(n,p)}X,\widehat{c}(k_n,p-),\widehat{c}(k_n,p+)\bigr)
-F(\Delta X_{T_p},c_{T_p-},c_{T_p}) \bigr)\\
&&\qquad\hspace*{0pt}{}- \bigl(F'_2(\Delta X_{T_p},c_{T_p-},c_{T_p})
\kappa(k_n,p-)\\
&&\qquad\hspace*{15pt}{} + F'_3(\Delta X_{T_p},c_{T_p-},c_{T_p})\kappa(k_n,p+) \bigr)
\stackrel{\mathbb{P}}{\longrightarrow}0.
\end{eqnarray*}

The sequences $\kappa(k_n,p\pm)^n$ are bounded in probability and
$\Delta X_{T_p}\in R$ a.s., so (\ref{Z-101}) and Taylor's formula yield
\begin{eqnarray*}
&&\hspace*{-2pt}\sqrt{k_n} \bigl(F(\Delta X_{T_p},\widehat{c}(k_n,p-),\widehat{c}(k_n,p+))-
F(\Delta X_{T_p},c_{T_p-},c_{T_p}) \bigr)\\
&&\hspace*{-2pt}\qquad{} - F'_2(\Delta X_{T_p},c_{T_p-},c_{T_p})
\kappa(k_n,p-)-F'_3(\Delta X_{T_p},c_{T_p-},c_{T_p})\kappa(k_n,p+)
\stackrel{\mathbb{P}}{\longrightarrow}0.
\end{eqnarray*}
So in fact it is enough to prove that
%
%
\begin{eqnarray}\label{PR-33}
&&\sqrt{k_n} \bigl(F\bigl(\Delta^n_{i(n,p)}X,\widehat{c}(k_n,p-),\widehat
{c}(k_n,p+)\bigr)\nonumber\\[-8pt]\\[-8pt]
&&\qquad{}-F(\Delta X_{T_p},\widehat{c}(k_n,p-),\widehat{c}(k_n,p+)) \bigr)
\stackrel{\mathbb{P}}{\longrightarrow}0.\nonumber
\end{eqnarray}
Since $\Delta X_{T_p}\in R$ a.s. and the two sequences $\widehat{c}(k_n,p-)$
and $\widehat{c}(k_n,p+)$ are tight in $(0,\infty)$, the first part
of (\ref{Z-101})
yields that (\ref{PR-33}) will hold if
$\sqrt{k_n} |\Delta^n_{i(n,p)}X-\Delta X_{T_p}|\stackrel{\mathbb
{P}}{\longrightarrow}0$. Therefore (\ref{PR-33})
follows from the facts that $k_n\Delta_n\to0$ and that the sequence
$\frac1{\sqrt{\Delta_n}} |\Delta^n_{i(n,p)}X-\Delta X_{T_p}|$ is
bounded in probability,
the latter coming, for example, from Lemma 8.5 of \cite{JSEMSTAT}. This
ends the proof of (\ref{PR-33}), hence of (\ref{PR-322}).

\textit{Step} 3. Here we prove (i). Suppose first that $F(x,y,z)=0$ for
$|x|\leq\varepsilon$ for some $\varepsilon>0$, and take
$m>2/\varepsilon$. As in the previous
theorem we then have $U(F)=\widetilde{U}(m)$ and $\mathcal U=\mathcal U(m)$
and $\mathcal U'=\mathcal U'(m)$, whereas $U(F,k_n)_s=\widetilde
{U}(m)_s$ for all $s\leq t$
on a set $\Omega^n_t$ having $\mathbb{P}(\Omega^n_t)\to1$. The result
follows from (\ref{PR-322}).

Next we assume $r=0$. Again as in the previous proof, we argue with
$m=\infty$: we have $U(F)=\widetilde{U}(\infty)$ and $\mathcal
U=\mathcal U(\infty)$ and
$\mathcal U'=\mathcal U'(\infty)$, whereas $U(F,k_n)_s=\widetilde
{U}(\infty)_s$ for all
$s\leq t$
on a set $\Omega'^n_t$ having $\mathbb{P}(\Omega'^n_t)\to1$. Then
the result
follows as before.

\textit{Step} 4. Now we assume $r>0$. By (\ref{Z-8}) and the
boundedness of $c_t$, we have
\[
\widetilde{\mathbb{E}}\bigl(|\mathcal U_t-\mathcal U(m)_t|^2\mid\mathcal
F\bigr) \leq
K\sum_{s\leq t}|\Delta X_s|^{2r} 1_{\{|\Delta X_s|\leq1/m\}}
\]
as soon as $m\geq1/\varepsilon$. This goes to $0$ a.s. as $m\to
\infty$ because
of Assumption \ref{asssumHr}, and it follows that $\mathcal U(m)\stackrel{\mathrm
{u.c.p.}}{\longrightarrow}\mathcal U$ (convergence in
probability, locally uniformly in time). In the same way, we have
$\mathcal U'(m)\stackrel{\mathrm{u.c.p.}}{\longrightarrow}\mathcal
U'$. Therefore, it remains to prove that for all
\mbox{$t,\eta>0$},
%
%
\begin{equation}\label{PR-324}
\lim_{m\to\infty} \limsup_{n\to\infty} \mathbb{P}\bigl(
\sqrt{k_n} |\widehat{U}^n(m)_t|>\eta\bigr) = 0
\end{equation}
and the same for $\widehat{U}'^n(m)$. We will prove (\ref{PR-324}) only.
Observe that, with the simplifying notation $c^n_i=c_{i\Delta_n}$ and
$c'^n_i=c_{(i-1)\Delta_n}$, we have $\widehat{U}^n(m)=\sum
_{j=1}^2\overline{V}(m,j)^n+
\sum_{j=1}^3V(m,j)^n$, where $V(m,j)^n_t=\sum_{i\in J(n,m,t)}\zeta(m,j)^n_i$
and, with $J'(n,m,t)=\{i\dvtx 1\leq i\leq[t/\Delta_n]\}\cap J(n,m,t)^c$,
\begin{eqnarray*}
\overline{V}(m,1)^n_t&=&-\sum_{i\in J'(n,m,t)}
\sum_{s\in I(n,i)}
F(\Delta\overline{Y}(m)_s,c_{s-},c_s),\\
\overline{V}(m,2)^n_t&=&-\sum_{0<s\leq k_n\Delta_n}F(\Delta
X_s,c_{s-},c_s)\\
&&{}
-\sum_{[t/\Delta_n]-k_n)\Delta_n<s\leq t}F(\Delta X_s,c_{s-},c_s),\\
\zeta(m,1)^n_i&=&
\bigl(F(\Delta^n_i\overline{Y}(m),\widehat{c}(k_n)_{i-k_n-1},\widehat
{c}(k_n)_i)\\
&&\hspace*{47.74pt}{}
-F(\Delta^n_i\overline{Y}(m),c'^n_i,c^n_i) \bigr)1_{\{\Delta^n_i\overline
{Y}(m)|>u_n\}},\\
\zeta(m,2)^n_i&=&
F(\Delta^n_i\overline{Y}(m),c'^n_i,c^n_i) 1_{\{\Delta^n_i\overline
{Y}(m)|>u_n\}}\\
&&{}
-\sum_{s\in I(n,i)}F(\Delta\overline{Y}(m)_s,c'^n_i,c^n_i),\\
\zeta(m,3)^n_i&=&\sum_{s\in I(n,i)} \bigl(
F(\Delta\overline{Y}(m)_s,c'^n_i,c^n_i)-F(\Delta\overline
{Y}(m)_s,c_{s-},c_s) \bigr).
\end{eqnarray*}
In view of (\ref{PR-22}) we are thus left to prove the existence
of sets $\Omega(n,m,t,j)$ and $\overline{\Omega}(n,m,t,j)$
satisfying for all
$m\geq2/\varepsilon$,
%
%
\begin{equation}\label{PR-325}
\lim_{n\to\infty} \mathbb{P}(\Omega(n,m,t,j)) = 1,\qquad
\lim_{n\to\infty} \mathbb{P}(\overline{\Omega}(n,m,t,j)) = 1,
\end{equation}
such that, for $j=1,2$ and $j=1,2,3$, respectively,
%
%
\begin{eqnarray}\label{PR-3270}
\lim_{m\to\infty} \limsup_{n\to\infty} \sqrt{k_n} \mathbb{E}
\bigl(1_{\overline{\Omega}(n,m,t,j)}
|\overline{V}(m,j)^n_t| \bigr) &=& 0,
\\
%
\label{PR-327}
\lim_{m\to\infty} \limsup_{n\to\infty} \sqrt{k_n} \mathbb{E}
\Biggl(1_{\Omega(n,m,t,j)}
\sum_{i=1}^{[t/\Delta_n]}|\zeta(m,j)^n_i| \Biggr) &=& 0.
\end{eqnarray}

\textit{Step} 5. In this step we prove (\ref{PR-3270}). In view
of the second part of (\ref{Z-101}) and of $F(0,y,z)=0$ and (\ref{PR-1})
we have when $m>1/\varepsilon$,
\[
{\sum_{s\in I(n,i)}}|F(\Delta\overline{Y}(m)_s,c_{s-},c_s)| \leq
a(n,i) =
{K\sum_{s\in I(n,i)}}|\Delta\overline{Y}(m)_s|^p.
\]
Moreover, we have the following estimate, for all $i$ possibly random but
$\mathcal F_0^{(m)}$-measurable:
%
%
\begin{equation}\label{Z-1040}
\mathbb{E}\bigl(a(n,i)\mid\mathcal F_0^{(m)}\bigr) \leq K\Delta_n\int
_{A_m}\gamma(z)^p \lambda(dz)
\leq K\Delta_n\gamma_m.
\end{equation}
Since $k_n\Delta_n\to0$ the set $\overline{\Omega}(n,m,t,2)=\{
D_m\cap[0,k_n\Delta_n]=
\varnothing,D_m\cap[t-(k_n+1)\Delta_n,t]=\varnothing\}$ satisfies
(\ref{PR-325}), and on this set we have $|\overline{V}(m,2)^n_t|\leq
\sum_{i=1}^{k_n}a(n,i)+\sum_{i=[t/\Delta_n]-k_n}^{[t/\Delta_n]+1}a(n,i)$.
Then (\ref{PR-3270}) for $j=2$ readily follows from (\ref{Z-1040})
and the property $k_n^{3/2}\Delta_n\leq K$ [see (\ref{Z-103})].

Now we consider\vspace*{-2pt} the case $j=1$. We have $|\overline{V}(m,1)^n_t|\leq
\sum_{i\in J'(n,m,t)}^{k_n}a(n,i)$. The successive integers in
$J'(n,m,t)$ are $\mathcal F_0^{(m)}$-measurable, and the number of them is
a Poisson variable independent of the $a(n,i)$'s and with
some parameter $\alpha(m,t)$ (exploding with $m$). Then
$\mathbb{E}(|\overline{V}(m,1)^n_t|)\leq K\alpha(m,t)\Delta_n$, and
(\ref{PR-3270}) for
$j=1$ holds with $\overline{\Omega}(n,m,t,1)=\Omega$.

\textit{Step} 6. In this step we prove (\ref{PR-327}) for $j=1$.
The sets
%
%
\begin{equation}\label{Z-120}
\Omega(n,m,t,1) = \bigcap_{i\leq[t/\Delta_n]}
\{|\Delta^n_i\overline{Y}(m)|\leq2/m, |\Delta^n_iX'(m)|\leq u_n/2\},
\end{equation}
satisfy the first part of (\ref{PR-325}) because
$|\Delta\overline{Y}(m)_s|\leq1/m$ and $\mathbb{P}(|\Delta
^n_iX'(m)|>u_n/2)\leq
K_m\Delta_n^2$ [use (\ref{PR-4}) for this].
When $m\geq2/\varepsilon$, (\ref{Z-101}) yields that\break $|\zeta
(m,1)^n_i|\leq
\zeta(m,4)^n_i$ on the set $\Omega(n,m,t,1)$ and for all $i\leq
[t/\Delta
_n]$ where
\[
\zeta(m,4)^n_i = K|\Delta^n_iX''(m)|^r \bigl(|\widehat{c}(k_n)_{i-k_n-1}-c'^n_i|
+|\widehat{c}(k_n)_i-c_i^n| \bigr).
\]
Then it remains to prove that (\ref{PR-327}) holds for $j=4$ and
$\Omega(n,m,t,4)=\Omega$.

Apply (\ref{PR-4030}) with $m=0$ and $S=(i-1)\Delta_n$ or $S=i\Delta_n$
[so $\Omega(0,n,S,i)_\pm=\Omega$] to get
%
%
\begin{equation}\label{Z-105}\qquad
\mathbb{E}\bigl(|\widehat{c}(k_n)_{i-k_n-1}-c_i'^n|\bigr) \leq\frac
K{\sqrt{k_n}},\qquad
\mathbb{E}\bigl(|\widehat{c}(k_n)_i-c_i^n|\mid\mathcal F_{i\Delta_n}\bigr)
\leq\frac K{\sqrt{k_n}}.
\end{equation}
Moreover (\ref{PR-4}) gives $\mathbb{E}(|\Delta^n_iX''(m)|^r\mid
\mathcal F_{(i-1)\Delta
_n})\leq
K\Delta_n\gamma_m$. Then by successive conditioning we obtain
$\mathbb{E}(\zeta
(m,4)^n_i)\leq
K\Delta_n\gamma_m/\sqrt{k_n}$. Since $\gamma_m\to0$ as $m\to
\infty$ we
deduce (\ref{PR-327}).

\textit{Step} 7. Now we prove (\ref{PR-327}) for $j=3$ with
$\Omega(n,m,t,3)=\Omega$.
We suppose that $m\geq1/\varepsilon$, so $|\Delta\overline
{Y}(m)_s|\leq\varepsilon$ and (\ref{Z-101})
yields that $|\zeta(m,3)^n_i|\leq K(\zeta(m,5)^n_i+\zeta(m,6)^n_i)$ where
\begin{eqnarray*}
\zeta(m,5)^n_i&=&{\sum_{s\in I(n,i)}}|\Delta\overline{Y}(m)_s|^r
|c_{s-}-c_i'^n|,\\
\zeta(m,6)^n_i&=&{\sum_{s\in I(n,i)}}|\Delta\overline{Y}(m)_s|^r
|c_i^n-c_s|.
\end{eqnarray*}
So it is enough to prove (\ref{PR-327}) for $j=5,6$.
The case $j=5$ is simple: the process $|c_{s-}-c_i'^n|
1_{s>(i-1)\Delta_n}$ is predictable; hence
\begin{eqnarray*}
\mathbb{E}(\zeta(m,5)^n_i)&=&\mathbb{E}\biggl(\int_{I(n,i)}\int_{A_m}
|c_{s-}-c_i'^n| |\delta(s,z)|^r \mu(ds,dz) \biggr)\\
&=&\mathbb{E}\biggl(\int_{I(n,i)}\,ds\int_{A_m}
|c_{s-}-c_i'^n| |\delta(s,z)|^r \lambda(dz) \biggr)\\
&\leq&\gamma_m\int_{I(n,i)}\mathbb{E}(|c_{s-}-c_i'^n|)\,ds
\leq K\Delta_n^{1+v}\gamma_m,
\end{eqnarray*}
where the last inequality comes from (\ref{S201}) with $m=0$
and $R=(i-1)\Delta_n$. Then (\ref{PR-327}) for $j=5$ follows because
$\Delta_n^v\sqrt{k_n}\to0$ by (\ref{Z-103}).

For $j=6$ we use again (\ref{S201}) with $m=0$ and $R=T_p$ below
to get
\begin{eqnarray*}
\mathbb{E}(\zeta(m,6)^n_i)&=&\sum_{p\geq1}\mathbb{E}\bigl(|\Delta
\overline{Y}(m)_{T_p}|^r
|c_i^n-c_{T_p}| 1_{I(n,i)}(T_p) \bigr)\\
&\leq&K\Delta_n^v\sum_{p\geq1}\mathbb{E}\bigl(|\Delta\overline{Y}(m)_{T_p}|^r
1_{I(n,i)}(T_p) \bigr)\\
&\leq&K\Delta_n^v\mathbb{E}\biggl(\sum_{s\in I(n,i)}|\Delta\overline
{Y}(m)_s|^r \biggr)
\leq K\Delta_n^{1+v}\gamma_m
\end{eqnarray*}
and we conclude as above.

\textit{Step} 8. Now we start proving (\ref{PR-327}) for $j=2$. Set
\begin{eqnarray*}
\zeta(m,7)^n_i&=&
F(\Delta^n_i\overline{Y}(m),c'^n_i,c^n_i) 1_{\{\Delta^n_i\overline
{Y}(m)|>u_n\}}\\
&&{}
-\sum_{s\in I(n,i)}F(\Delta\overline{Y}(m)_s,c'^n_i,c^n_i)
1_{\{\Delta\overline{Y}(m)_s|>u_n\}}.
\end{eqnarray*}
If $m\geq1/\varepsilon$, we
deduce from (\ref{Z-101}) and the boundedness of $c_t$ that
\[
|\zeta(m,2)^n_i-\zeta(m,7)^n_i| \leq {K\sum_{s\in I(n,i)}}
|\Delta\overline{Y}(m)_s|^p 1_{\{|\Delta\overline{Y}(m)_s|\leq u_n\}}.
\]
Therefore
\[
\mathbb{E}\bigl(|\zeta(m,2)^n_i-\zeta(m,7)^n_i|\bigr) \leq K\Delta_n
\int_{\{z\dvtx \gamma(z)\leq u_n\}}\gamma(z)^p\lambda(dz) \leq
K\Delta_n^{1+\varpi(p-r)}\gamma_m.
\]
Taking (\ref{Z-103}) into consideration, we deduce that
\[
\lim_{n\to\infty} \sqrt{k_n} \mathbb{E}\Biggl({\sum_{i=1}^{[t/\Delta_n]}}
|\zeta(m,2)^n_i-\zeta(m,7)^n_i| \Biggr) = 0
\]
and thus we are left to prove (\ref{PR-327}) for $j=7$.

\textit{Step} 9. In this auxiliary step we fix $m>2/\varepsilon$, and
also some $l\in(1,1/2r\varpi)$ [this is possible by (\ref{Z-103})].
We write $q_n=[(u_n)^{-l}]$ and
we suppose that $n$ is big enough for having $1/q_n<u_n<1/m$.
We complement notation (\ref{PR-23}) with
%
%
\begin{eqnarray}\label{Z-115}\hspace*{28pt}
A'_n &=& A_m\cap(A_{q_n})^c,\qquad
Y^n_t=\int_0^t\int_{A'_n}\delta(s,z)\mu(ds,dz),\nonumber\\
b^n_t &=& \cases{
\displaystyle-\int_{A'_n}\delta(t,z)\lambda(dz),&\quad if $r>1$,\cr
0, &\quad if $r\leq1$,}\qquad
B^n_t=\int_0^tb^n_s \,ds,\nonumber\\
\overline{Y}{}^n &=& \overline{Y}(m)-Y^n = X'(q_n)+X''(q_n)+B^n,\\
N^n_t &=& \mu([0,t]\times A'_n),\qquad
H(n,i)= \biggl\{|\Delta^n_i\overline{Y}{}^n|\leq\frac{u_n}2 \biggr\}\cap\{\Delta
^n_iN^n\leq
1 \}.\nonumber
\end{eqnarray}

First, $N^n$ is a Poisson process with parameter $\lambda(A'_n)\leq
K\gamma_mq_n^r$; hence
%
%
\begin{equation}\label{PR-51}
\mathbb{P}\bigl(\Delta^n_iN^n\geq2\mid\mathcal F^{(m)}_{(i-1)\Delta_n}\bigr)
\leq K\Delta
_n^{2-2rl\varpi}\gamma_m.
\end{equation}
Second, upon observing that $\Delta_nq_n^r\leq K$ (because $rl\varpi
\leq1$)
and $|b^n_t|\leq q_n^{r-1}\gamma_m$ when $r>1$ and $b^n_t=0$ if $r\leq
1$, that
%
%
\begin{equation}\label{PR-52}\quad
\iota\geq r \quad\Rightarrow\quad
\mathbb{E}\bigl(|\Delta^n_i\overline{Y}{}^n|^{\iota}\mid\mathcal
F_{(i-1)\Delta_n}^{(q_n)}\bigr) \leq K_{\iota
} \bigl(\Delta_n^{\iota/2}
+\Delta_n^{1+l\varpi(\iota-r)}\gamma_m \bigr).
\end{equation}
This applied with $\iota=\frac4{1-2\varpi}\vee\frac{1+lr\varpi
}{\varpi(l-1)}$
and Markov's inequality yield
%
%
\begin{equation}\label{PR-53}
\mathbb{P}(|\Delta^n_i\overline{Y}{}^n|>u_n/2) \leq K\Delta_n^2.
\end{equation}

Next, on the set $H(n,i)$, we have
$|\Delta^n_i\overline{Y}{}^n|\leq u_n/2$ and $|\Delta^n_iY^n|\leq
1/m$, and also
$|\Delta\overline{Y}(m)_s|\leq u_n$ for all $s\in I(n,i)$, except
when $\Delta^n_iN^n=1$
for a single value of $s$ for which $\Delta\overline{Y}(m)_s=\Delta
^n_iY^n$ (whose absolute
value may be smaller or greater than~$u_n$). In other words, on $H(n,i)$
we have
\begin{eqnarray*}
\zeta(m,7)^n_i&=& \bigl(
F(\Delta^n_iY^n+\Delta^n_i\overline{Y}{}^n,c'^n_i,c^n_i) 1_{\{|\Delta
^n_iY^n+\Delta^n_i\overline{Y}{}^n|>u_n\}}\\
&&\hspace*{57.1pt}{} - F(\Delta^n_iY^n,c'^n_i,c^n_i) 1_{\{|\Delta^n_iY^n|>u_n\}}
\bigr)\\
&&{}\times
1_{\{|\Delta^n_iY^n|\leq1/m, |\Delta^n_i\overline{Y}{}^n|\leq u_n/2\}}.
\end{eqnarray*}
The following estimate, when $u\in(0,1/m)$ and $y,z\in(0,M]$ for some
$M$ (this will be the bound of the process $c_t$) and $x,x'\in\mathbb
{R}$ with
$|x|\leq1/m$ and $|x'|\leq u/2$, is easy to prove, upon using (\ref{Z-101}):
\[
\bigl|F(x+x',y,z)1_{\{|x+x'|>u\}}-F(x,y,z)1_{\{|x|>u\}} \bigr|\leq
K \bigl(|x|^{p-1} |x'|+(|x|\wedge u)^{p} \bigr).
\]
Therefore, on the set $H(n,i)$ again we have
%
%
\begin{equation}\label{PR-54}
|\zeta(m,7)^n_i| \leq
K \bigl(|\Delta^n_iY^n|^{p-1}|\Delta^n_i\overline{Y}{}^n|+(|\Delta
^n_iY^n|\wedge u_n)^p \bigr).
\end{equation}

The process $Y^n$ satisfies the same estimate as $X''(m)$ in (\ref{PR-5}),
hence since $p\geq r$,
%
%
\begin{equation}\label{PR-55}\quad
\mathbb{E}\bigl((|\Delta^n_iY^n|\wedge u_n)^p\mid\mathcal
F^{(m)}_{(i-1)\Delta_n} \bigr)
\leq K\Delta_nu_n^{p-r}\gamma_m \leq K\Delta_n^{1+(p-r)\varpi
}\gamma_m.
\end{equation}
On the other hand, we can apply (\ref{PR-52}) with $\iota=2$ and
the Cauchy--Schwarz inequality to obtain\vspace*{-2pt} $\mathbb{E}(|\Delta
^n_i\overline{Y}{}^n|\mid\mathcal F
_{(i-1)\Delta_n}
^{(q_n)})\leq K\sqrt{\Delta_n}$.
We also have $|\Delta^n_iY^n|\leq\Delta^n_iG(A'_n)$ [see before
(\ref{PR-2031})
for this notation], and $\Delta^n_iG(A'_n)$ is
$\mathcal F_0^{(q_n)}$-measu\-rable. Therefore, in view of (\ref
{PR-2031}) applied
with the power $(p-1)\vee r$ and H\"older's inequality, and upon applying
$(r\vee1)(1-(r-1)^+l\varpi)\geq1$,
and with the notation $q=1\wedge\frac{p-1}r$, we see that
\begin{eqnarray*}
\mathbb{E}(|\Delta^n_iY^n|^{p-1}|\Delta^n_i\overline{Y}{}^n| )
&=&\mathbb{E}\bigl(|\Delta^n_iY^n|^{p-1} \mathbb{E}\bigl(|\Delta^n_i\overline{Y}{}^n|
\mid\mathcal F^{(q_n)}_{(i-1)\Delta_n} \bigr) \bigr)\\
&\leq& K\sqrt{\Delta_n}\mathbb{E}(|\Delta^n_iY^n|^{p-1} ) \leq
K\Delta_n^{1/2+q}
\gamma_m^q.
\end{eqnarray*}
Hence by (\ref{PR-54}) and (\ref{PR-55}), we deduce
%
%
\begin{equation}\label{PR-56}
\mathbb{E}\bigl(|\zeta(m,7)^n_i| 1_{H(n,i)} \bigr) \leq K\gamma_m^q
\bigl(\Delta_n^{1+(p-r)\varpi}+\Delta_n^{1/2+q} \bigr).
\end{equation}

\textit{Step} 10. Now we are ready to prove the result for
$j=7$. We take\break $\Omega(n,m,t,7)=\bigcap_{1\leq i\leq[t/\Delta_n]}H(n,i)$,
which by
(\ref{PR-51}) and (\ref{PR-53}) satisfies
\[
\mathbb{P}(\Omega(n,m,t,7)^c) \leq\sum_{i=1}^{[t/\Delta_n]}\mathbb
{P}(H(n,i)^c) \leq Kt\Delta
_n^{1-2rl\varpi},
\]
hence (\ref{PR-325}) because $2rl\varpi<1$.
Finally,
\[
\mathbb{E}\Biggl({1_{\Omega(n,m,t,7)}\sum_{i=1}^{[t/\Delta_n]}}|\zeta
(m,7)^n_i|\Biggr)\leq
\sum_{i=1}^{[t/\Delta_n]}\mathbb{E}\bigl(|\zeta(m,7)^n_i| 1_{H(n,i)}\bigr),
\]
so (\ref{PR-56}) shows that
(\ref{PR-325}) holds, provided the sequences
$\Delta_n^{(p-r)\varpi}\sqrt{k_n}$ and
$\Delta_n^{q-1/2}\sqrt{k_n}$ are bounded. These amount to having
$2(p-r)\geq\rho$ and $2q-1\geq\rho$, which are
implied by (\ref{Z-103}).

\subsection{\texorpdfstring{Proof of Theorem \protect\ref{TKP-3}.}{Proof of Theorem 3.3}}

\textit{Step} 1. We assume Assumptions \ref{asssumHr} and \ref{asssumKv} and (\ref{PR-1}).
Recalling (\ref{S1}) and (\ref{PR-0}), we set $\overline{\delta
}(t,z)=\break\delta
(t,z)
1_{\{\delta(t,z)\notin A\}\cup\{\widehat{\delta}(t,z)=0\}}$, and define
$\overline{X}$ by (\ref{S1}) with $\overline{\delta}$ instead of
$\delta$. This process
satisfies Assumption \ref{asssumHr} as well, and coincides with $X$ on the interval
$[0,t]$, in restriction to the set $\Omega^A_t$. Hence the variables
$U(F,k_n)_t$ and $U(F)_t$ and $\overline{\mathcal U}_t$ and $\overline
{\mathcal U}{}'_t$ are the same
on $\Omega^A_t$, whether computed using $X$ or $\overline{X}$.
So it is enough to prove the result for the process $\overline{X}$.
Or, in
other words, we can assume throughout that
%
%
\begin{equation}\label{PR-70}
\Delta X_s\in A\setminus\{0\} \quad\Rightarrow\quad\Delta\sigma_s=0 \mbox
{ identically.}
\end{equation}

We use the same arguments as in the previous proof, and the
same notation, except that the variable $\overline{\zeta}{}^n_p$ of
(\ref{PR-321})
should be replaced by
\begin{eqnarray*}
\overline{\zeta}{}^n_p &=& \tfrac12 \bigl(F''_{22}(\Delta X_{T_p},c_{T_p-},c_{T_p})
\kappa(k_n,p-)^2\\
&&\hspace*{8.90pt}{} + F''_{33}(\Delta X_{T_p},c_{T_p-},c_{T_p})\kappa
(k_n,p+)^2\\
&&\hspace*{8.90pt}{} + 2F''_{23}(\Delta X_{T_p},c_{T_p-},c_{T_p})\kappa(k_n,p+)
\kappa(k_n,p-) \bigr) 1_{\{\Delta X_{T_p}\neq0\}}
\end{eqnarray*}
and the same for $\overline{\zeta}{}'^n_p$ with $wk_n$ instead of $k_n$.

\textit{Step} 2. In this step we prove that
%
%
\begin{equation}\label{PR-60}
(k_n\widetilde{U}^n(m),
k_n\widetilde{U}'^n(m) ) \quad\stackrel{\mathcal L-\mathrm
{s}}{\Longrightarrow}\quad (\overline{\mathcal U}(m),\overline{\mathcal
U}{}'(m) ),
\end{equation}
where $\overline{\mathcal U}(m)$ and $\overline{\mathcal U}{}'(m)$ are
as in (\ref{Z-13}), except that
the sum
is taken over the $p\in\mathcal{T}_m$ only. By Proposition \ref
{PPR-1}, we have
\[
\biggl(\sum_{p\in\mathcal{T}_m}\overline{\zeta}{}^n_p 1_{\{T_p\leq t\}},
\sum_{p\in\mathcal{T}_m}\overline{\zeta}{}'^n_p 1_{\{T_p\leq t\}} \biggr)
\quad\stackrel{\mathcal L-\mathrm{s}}{\Longrightarrow}\quad
(\overline{\mathcal U}(m)_t,w\overline{\mathcal U}{}'(m)_t ),
\]
so proving (\ref{PR-322}) shows that for each $p\in\mathcal{T}_m$
and on each set $\{T_p\leq t\}$ we have
%
%
\begin{equation}\label{PR-61}
k_n \zeta^n_p-\overline{\zeta}{}^n_p \stackrel{\mathbb
{P}}{\longrightarrow}0,\qquad wk_n \zeta'^n_p-\overline{\zeta}{}
'^n_p \stackrel{\mathbb{P}}{\longrightarrow}0.
\end{equation}
We prove only the first property, which [like in Theorem \ref{TKP-2};
note that here $F(\Delta X_{T_p},c_{T_p-},c_{T_p})=0$ by
(\ref{Z-104}) and (\ref{PR-70})] amounts to the convergence of
\begin{eqnarray*}
&&k_nF\bigl(\Delta^n_{i(n,p)}X,\widehat{c}(k_n,p-),\widehat{c}(k_n,p+)\bigr)\\
&&\qquad{} - \tfrac12 \bigl(F''_{22}(\Delta X_{T_p},c_{T_p-},c_{T_p})
\kappa(k_n,p-)^2\\
&&\hspace*{22.1pt}\qquad{} + 2F''_{23}(\Delta
X_{T_p},c_{T_p-},c_{T_p})\kappa(k_n,p+)\kappa(k_n,p-)\\
&&\hspace*{67.6pt}\qquad{}
+ F''_{33}(\Delta X_{T_p},c_{T_p-},c_{T_p})\kappa(k_n,p+)^2 \bigr)
\end{eqnarray*}
to $0$ in probability. Upon using again (\ref{Z-104}) and (\ref{PR-70}),
we deduce from Taylor's formula and the tightness of the sequences
$\kappa(k_n,p\pm)$ that, on the set $\{\Delta X_{T_p}\in R\}$ which has
probability $1$, the variables
\begin{eqnarray*}
&&k_nF(\Delta X_{T_p},\widehat{c}(k_n,p-),\widehat{c}(k_n,p+))\\
&&\qquad{}-\tfrac12 \bigl(
F''_{22}(\Delta X_{T_p},c_{T_p},c_{T_p})\kappa(k_n,p-)^2\\
&&\qquad\hspace*{22.1pt}{} + 2F''_{23}(\Delta X_{T_p},c_{T_p},c_{T_p})\kappa(k_n,p-)
\kappa(k_n,p+)\\
&&\hspace*{90.3pt}{} + F''_{33}(\Delta X_{T_p},c_{T_p},c_{T_p})\kappa
(k_n,p+)^2 \bigr)
\end{eqnarray*}
go to $0$ in probability.
Hence the first part of (\ref{PR-61}) will follow if we show
\[
k_n \bigl( F\bigl(\Delta^n_{i(n,p)}X,\widehat{c}(k_n,p-),\widehat{c}(k_n,p+)\bigr)
- F(\Delta X_{T_p},\widehat{c}(k_n,p-),\widehat{c}(k_n,p+)) \bigr)
\stackrel{\mathbb{P}}{\longrightarrow}0.
\]
This is proved exactly as (\ref{PR-33}), except that here we use the
property $k_n\sqrt{\Delta_n}\to0$.

\textit{Step} 3. The proof of (i) follows from (\ref{PR-60}) in
exactly the same way as in Step~3 of the proof of Theorem \ref{TKP-2}.

\textit{Step} 4. Now we start proving (ii), so $r>0$. We can suppose that
$A$ contains a neighborhood of $0$; otherwise we are in the second situation
of case (i). Hence we may take $\varepsilon>0$ in (\ref{Z-102}) such
that also
$[-\varepsilon,\varepsilon]\subset A$. Similar to (\ref{Z-801}),
and by the boundedness of $c_t$ and (\ref{Z-102}), we have if $m\geq
1/\varepsilon$,
\[
\widetilde{\mathbb{E}}\bigl(|\overline{\mathcal U}_t-\overline{\mathcal
U}(m)_t|\mid\mathcal F\bigr) \leq
{K\sum_{s\leq t}}|\Delta X_s|^{r} 1_{\{|\Delta X_s|\leq1/m\}}.
\]
This goes to $0$ a.s. as $m\to\infty$ because
of Assumption \ref{asssumHr}, so $\overline{\mathcal U}(m)\stackrel{\mathrm
{u.c.p.}}{\longrightarrow}\overline{\mathcal U}$, and also
$\overline{\mathcal U}{}'(m)\stackrel{\mathrm{u.c.p.}}{\longrightarrow
}\overline{\mathcal U}{}'$. Then it remains to prove that for all
$t,\eta>0$,
%
%
\begin{equation}\label{PR-62}
\lim_{m\to\infty} \limsup_{n\to\infty} \mathbb{P}\bigl(
k_n |\widehat{U}^n(m)_t|>\eta\bigr) = 0
\end{equation}
and the same for $\widehat{U}'^n(m)$. We will prove (\ref{PR-62}) only.

Because of our assumptions
we have here $\widehat{U}^n(m)=\overline{U}{}^n(m)$. Then, in view of definition
(\ref{PR-25}), and since the sets $\Omega(n,m,t,1)$ of (\ref{Z-120})
satisfy (\ref{PR-325}), it is enough to prove that
%
%
\begin{equation}\label{PR-63}
\lim_{m\to\infty} \limsup_{n\to\infty} k_n \mathbb{E}\Biggl({1_{\Omega
(n,m,t,1)}
\sum_{i=1}^{[t/\Delta_n]}}|\zeta(m,1)^n_i| \Biggr) = 0,
\end{equation}
where
\[
\zeta(m,1)^n_i=F(\Delta^n_i\overline{Y}(m),\widehat
{c}(k_n)_{i-k_n-1},\widehat{c}(k_n)_i)
1_{\{\Delta^n_i\overline{Y}(m)|>u_n\}}.
\]
On $\Omega(n,m,t,1)$, when $m>2/\varepsilon$, for all $i\leq
[t/\Delta_n]$
we have $|\Delta^n_i\overline{Y}(m)|\leq\varepsilon$ and also
$|\Delta^n_i\overline{Y}(m)|\leq2|\Delta^n_iX''(m)|$
when further $|\Delta^n_i\overline{Y}(m)|>u_n$. Then, using (\ref
{Z-102}) and
a Taylor expansion around $(\Delta^n_i\overline{Y}(m),c_{i\Delta
_n},c_{i\Delta_n})$,
and since $F(x,y,y)=F'_2(x,y,y)=F'_3(x,y,y)=0$ for all $x,y$, we
see that
\[
|\zeta(m,1)^n_i|\leq K \bigl(\zeta(m,2)^n_i+\zeta(m,3)^n_i \bigr)\qquad
\mbox{on $\Omega(n,m,t,1)$ and for $i\leq[t/\Delta_n]$},
\]
where
\begin{eqnarray*}
\zeta(m,2)^n_i &=& |\Delta^n_iX''(m)|^r \bigl(\bigl|\widehat
{c}(k_n)_{i-k_n-1}-c_{(i-1)\Delta_n}\bigr|^2
+|\widehat{c}(k_n)_i-c_{i\Delta_n}|^2 \bigr),\\
\zeta(m,3)^n_i &=& |\Delta^n_iX''(m)|^r |\Delta^n_ic|^2.
\end{eqnarray*}
Hence we are left to prove that, for $j=2,3$, we have
%
%
\begin{equation}\label{PR-65}
\lim_{m\to\infty} \limsup_n k_n \mathbb{E}\Biggl(\sum_{i=1}^{[t/\Delta
_n]}\zeta(m,j)^n_i \Biggr) = 0.
\end{equation}

\textit{Step} 5. On the one hand, successive conditioning, plus the
third estimate in (\ref{PR-4}) with $p=r$, plus (\ref{PR-4030}) with
$m=0$ and $q=2$, yield $\mathbb{E}(\zeta(m,2)^n_i)\leq K\Delta
_n\gamma_m/k_n$.
Then (\ref{PR-65}) for $j=2$ follows. For $j=3$
we will prove the stronger statement, for $m$ large enough,
%
%
\begin{equation}\label{PR-66}
\lim_n k_n \mathbb{E}\Biggl(\sum_{i=1}^{[t/\Delta_n]}\zeta(m,j)^n_i \Biggr) = 0.
\end{equation}
Therefore, we fix $m\geq2/\varepsilon$ below.

First, suppose that $r\leq1$. Then $X''(m)_t=\sum_{s\leq t}
\Delta X''(m)_s$, and since $|x+x'|^r\leq|x|^r+|x'|^r$ and
$c_s=c_{s-}$ when $\Delta X(m)''_s\neq0$ [recall $m\geq2/\varepsilon
$ and
(\ref{PR-70})], we have $\zeta(m,3)^n_i\leq\zeta(m,4)^n_i$, where
\[
\zeta(m,4)^n_i=\sum_{s\in I(n,i)}|\Delta\overline{Y}(m)_s|^r \bigl|c_{s-}-
c_{(i-1)\Delta_n}\bigr|^2
+\sum_{s\in I(n,i)}|\Delta\overline{Y}(m)_s|^r |c_{i\Delta_n}-c_s|^2.
\]
Then exactly as in Step 7 of Theorem \ref{TKP-2}, and
using (\ref{S201}) with $p=2$ instead of $p=1$, we obtain
$\mathbb{E}(\zeta(m,3)^n_i)\leq K\Delta_n^{1+(2v)\wedge1}$. Then
(\ref{PR-66})
holds for $j=4$, hence for $j=3$, by (\ref{Z-113}).

It remains to consider the case $r>1$.
We take $l=1/r\varpi$, and we use the notation $q_n=[(u_n)^{-l}]$
and (\ref{Z-115}), which we complement as follows:
\[
Z(5)^n = B^n,\qquad Z(6)^n = X''(q_n),\qquad Z(7)^n = Y^n,
\]
so $X''(m)=\sum_{j=5}^7Z(j)^n$, and we
associate the variables
\[
\zeta(m,j)^n_i = |\Delta^n_iZ(j)|^r |\Delta^n_ic|^2.
\]
It is thus enough to prove (\ref{PR-66}) when $j=5,6,7$.
First, we have $|\Delta^n_iZ(5)^n|\leq K\Delta_n^{1-l\varpi
(r-1)}\gamma_m$, and thus
by (\ref{S201}) we get $\mathbb{E}(\zeta(m,5)^n_i)\leq K\Delta
_n^{r-(r-1)rl\varpi
+(2v)\wedge1}$, which equals $K\Delta_n^{1+(2v)\wedge1}$,
and (\ref{PR-66}) for $j=5$ holds by (\ref{Z-113}).
Next, (\ref{PR-4}) applied with $q_n$ instead of $m$ implies that
for any $p\geq2$ we have $\mathbb{E}(|\Delta^n_iZ(6)|^p)\leq
K_p\Delta_n^{p/r}$ (use
again $rl\varpi=1$). Then by (\ref{S201}) and H\"older's inequality we
see that $\mathbb{E}(\zeta(m,6)^n_i)\leq K_{\theta}\Delta
_n^{1+(2v)\wedge\theta}$
for any $\theta\in(1/2,1)$. Then again, upon taking $\theta$ close
to $1$,
we have (\ref{PR-66}) for $j=6$.

Finally, we set $Y(n,i)_t=\sum_{(i-1)\Delta_n<s\leq t}
|\Delta Y^n_s|$ for $t\in I(n,i)$. Observe that
\begin{eqnarray*}
|\Delta^n_iZ(7)^n|^r
&\leq&Y(n,i)_{\Delta_n}^r =
\sum_{s\in I(n,i)} \bigl(\bigl(Y(n,i)_{s-}+|\Delta
Y^n_s|\bigr)^r-Y(n,i)_{s-}^r \bigr)\\
&\leq& K\sum_{s\in I(n,i)}
\bigl(|\Delta Y^n_s|^r+Y(n,i)^{r-1}_{s-}|\Delta Y^n_s| \bigr).
\end{eqnarray*}
Since $|\Delta Y^n|\leq|\Delta\overline{Y}(m)|$, it follows that
$\zeta(m,7)^n_i\leq
K(\zeta(m,4)^n_i+\zeta(m,8)^n_i
+\zeta(m,9)^n_i)$, where $\zeta(m,4)^n_i$ is as in the case $r\leq1$ and
\begin{eqnarray*}
\zeta(m,8)^n_i &=& \sum_{s\in I(n,i)}Y(n,i)^{r-1}_{s-}
|\Delta Y^n_s| \bigl|c_{s-}-c_{(i-1)\Delta_n}\bigr|^2,\\
\zeta(m,9)^n_i &=& \sum_{s\in I(n,i)}Y(n,i)^{r-1}_{s-}
|\Delta Y^n_s| |c_{i\Delta_n}-c_s|^2.
\end{eqnarray*}
We have seen that (\ref{PR-66}) is satisfied for $j=4$ (this is
irrespective of the value of~$r$). For proving it for
$j=8$ and $j=9$ we use the same argument as in Step 7 of Theorem
\ref{TKP-2} again, thus getting
\begin{eqnarray*}
\mathbb{E}(\zeta(m,8)^n_i)&\leq&\gamma_m \int_{I(n,i)}
\mathbb{E}\bigl(Y(n,i)^{r-1}_{s-} \bigl|c_{s-}-c_{(i-1)\Delta_n}\bigr|^2 \bigr) \,ds,\\
\mathbb{E}(\zeta(m,9)^n_i)&\leq&K\Delta_n^{(2v)\wedge1}
\mathbb{E}\biggl(\sum_{s\in I(n,i)}Y(n,i)^{r-1}_{s-} |\Delta Y^n_s| \biggr)\\
&\leq&K\Delta_n^{(2v)\wedge1} \mathbb{E}\Bigl(\sup_{s\leq i\Delta
_n}(Y(n,i)_s)^r \Bigr).
\end{eqnarray*}

Note that $Y(n,i)$ has the same structure as $X''(q_n)$ does in case
$r\leq1$, so although $r>1$ here we have, as in the first part of
the third estimate in (\ref{PR-4}),
\begin{eqnarray*}
&&p\geq r \quad\Rightarrow\quad\mathbb{E}\Bigl(\sup_{s\leq i\Delta_n}(Y(n,i)_s)^p \Bigr)
\leq K_p \bigl(\Delta_n^{1+(p-r)l\varpi}+\Delta_n^{p+r(p-1)l\varpi} \bigr)\\
&&\hphantom{p\geq r \quad\Rightarrow\quad\mathbb{E}\Bigl(\sup_{s\leq i\Delta_n}(Y(n,i)_s)^p \Bigr)}
\leq K_p\Delta_n^{p/r}.
\end{eqnarray*}
Applying (\ref{S201}) and H\"older's inequality yields
$\mathbb{E}(\zeta(m,8)^n_i)\leq K_{\theta}\Delta_n^{r+(2v)\wedge
\theta}$
for any $\theta\in(1/2,1)$, whereas obviously
$\mathbb{E}(\zeta(m,9)^n_i)\leq K_{\theta}\Delta_n^{1+2v}$. Then
(\ref{PR-66})
holds for
$j=8$ and $j=9$.

\subsection{\texorpdfstring{Proof of the results on the
tests.}{Proof of the results on the tests}}

\mbox{}

\begin{pf*}{Proof of Theorem \protect\ref{TTE-1}} Theorems
\ref{TKP-1} and \ref{TKP-3} yield that, in restriction to
$\Omega_T^{(A,d)}$, the variables $k_nU(F,k_n)_T/U(G,k_n)_T$
converge stably to a positive variable $\mathcal V$ which,
conditionally on
$\mathcal F$,
has mean $1$. Hence if $H\subset\Omega_T^{(A,d)}$ and with $C_n$ given
by (\ref{TE-5}), we have $\limsup_n \mathbb{P}(C_n\cap H)\leq
\widetilde{\mathbb{P}}(H\cap
\{\mathcal V\geq1/\alpha\})$, which is smaller than $\alpha\mathbb
{P}(H)$ because
$\widetilde{\mathbb{E}}(\mathcal V\mid\mathcal F)=1$, and the result
for the asymptotic level follows. Since $k_nU(F,k_n)_T/U(G,k_n)_T
\stackrel{\mathbb{P}}{\longrightarrow}\infty$ on the set $\Omega
_T^{(A,j)}$ by Theorem \ref{TKP-1},
the asymptotic power is clearly $1$.
\end{pf*}
\begin{pf*}{Proof of Theorem \protect\ref{TTE-2}}
We will be very sketchy here.
By localization we may assume (\ref{PR-1}).

First, we can suppose that the simulated variables $V^\pm_i(j)$ are
defined on our auxiliary space\vspace*{1pt} $(\Omega',\mathcal
F',\mathbb{P}')$, so that the $\overline{\mathcal U}(n,j)$'s are
defined on the extension $(\widetilde{\Omega},\widetilde {\mathcal
F},\widetilde{\mathbb{P}})$. Then we can reproduce the proof of Theorem
4.4 of \cite{JT} to obtain that, if
$Z_n\stackrel{\mathbb{P}}{\longrightarrow}Z$ are $\mathcal
F$-measurable variables, we have
%
%
\begin{equation}\label{F1}
\widetilde{\mathbb{P}}\bigl(\overline{\mathcal U}(n,1)>Z_n\mid\mathcal
F\bigr) \stackrel{\mathbb{P}}{\longrightarrow}\widetilde{\mathbb
{P}}(\overline{\mathcal U}_T>Z\mid\mathcal F).
\end{equation}
The only slightly different point is that we need here
$\mathbb{E}((\widehat{c}(k_n)_i)^2\mid\mathcal F_{(i-1)\Delta
_n})\leq K$. This does not follow
from (\ref{PR-402}), but it does from (\ref{PR-4030}) applied with
$q=2$, because by hypothesis (\ref{PR-4031}) holds.

Then, using (\ref{F1}) and that $k_nU(F,k_n)_T\stackrel{\mathcal
L-(s)}{\longrightarrow} \overline{\mathcal U}_T$ on the set
$\Omega_T^{(A,d)}$, we can reproduce the proof of
Theorem 5.1, Part (c), of \cite{JT}, and we obtain the claim about
the asymptotic level. In the course of this proof it is also shown that
$\mathcal F$-conditionally the variables $\overline{\mathcal
U}_{(|N_n\alpha])}$ converge in
law to the unique variable $Z(\alpha)$ such that $\widetilde{\mathbb
{P}}(\overline{\mathcal U}_T>Z(\alpha)
\mid\mathcal F)=\alpha$, from which $\overline{\mathcal
U}_{(|N_n\alpha])}\stackrel{\mathbb{P}}{\longrightarrow}Z(\alpha)$ follows.

Finally $k_nU(F,k_n)_T\stackrel{\mathbb{P}}{\longrightarrow}\infty$
on $\Omega_T^{(A,j)}$. This
and $\overline{\mathcal U}_{(|N_n\alpha])}\stackrel{\mathbb
{P}}{\longrightarrow}Z(\alpha)$, yields that
$\widetilde{\mathbb{P}}(C_n\cap\Omega_T^{(A,j)})\to\mathbb
{P}(\Omega_T^{(A,j)}$. Hence the
asymptotic power equals $1$.
\end{pf*}
\begin{pf*}{Proof of Theorem \protect\ref{TTE-3}} The proof is the
same as
for Theorem \ref{TTE-1}, with the following changes: we now have
$\mathbb{P}(C_n\cap H)\to\alpha\mathbb{P}(H)$ because\break
$k_nU(F,k_n)_T$ converges
stably in
law on $\Omega_T^{(A,d)}$ to a chi-square variable with $N_T$ degrees of
freedom, independent of $\mathcal F$, and $N^n_T=N_T$ for $n$ large enough.
This gives\vspace*{-1pt} that the asymptotic level is\vspace*{1pt}
$\alpha$, and for the asymptotic
power we use the fact that $k_nU(F,k_n)_T\stackrel{\mathbb
{P}}{\longrightarrow}\infty$ and $N_T<\infty$
on the set $\Omega_T^{(A,j)}$.
\end{pf*}
\begin{pf*}{Proof of Theorem \protect\ref{TTE-4}}
The result readily follows
from the stable convergence in law of $(S_n-1)/\sqrt{V_n}$ to a
standard normal.
\end{pf*}
\begin{pf*}{Proof of Theorem \protect\ref{TTE-5}}
Since $V'_n=V_n$ for
all $n$ large enough, on the set $\Omega_T^{(A,j)}$, only the claim about
the power needs a proof. Now, $V'_n\to0$, and we have the
second part of (\ref{TE-12}) on $\Omega_T^{(A,d)}$; that the asymptotic
power equals $1$ is now obvious.
\end{pf*}

\section*{\texorpdfstring{Acknowledgments.}{Acknowledgments}}
We would like to thank Tim Bollerslev for providing us with the
high-frequency data and an anonymous referee for careful reading and
constructive comments on the paper.


%
\printaddresses

\end{document}